\documentclass[api,prb,letterpaper,twocolumn,groupedaddress,footinbib,amssymb,amsmath]{revtex4}

\usepackage{graphicx}
\usepackage{braket}
\usepackage{color}

\newcommand{\up}{\ensuremath{\uparrow}}
\newcommand{\down}{\ensuremath{\downarrow}}
\newcommand{\cmmc}{$\rm{Co_{\infty}}$-$\rm{Mn}$-$\rm{Mn}$-$\rm{Co_{\infty}}$~}
\newcommand{\SMN}{\ensuremath{S_{\rm{Mn}}}}
\newcommand{\SCO}{\ensuremath{S_{\rm{Co}}}}

\begin{document}

\title{Conductance fingerprints of non-collinear magnetic states in single atom contacts:
\newline a first-principles Wannier functions study}

\author{Bj\"orn Hardrat$^{1}$}
\author{Frank Freimuth$^{2}$}
\author{Stefan Heinze$^{1}$}
\author{Yuriy Mokrousov$^{2}$}
\email[corresp.\ author: ]{y.mokrousov@fz-juelich.de}

\affiliation{$^{1}$Institut f\"{u}r Theoretische Physik und
 Astrophysik, Christian-Albrechts-Universit\"{a}t zu Kiel,
 Leibnizstrasse~15, D-24098 Kiel, Germany}
\affiliation{$^{2}$Peter Gr\"{u}nberg Institut and Institute for Advanced Simulation,
Forschungszentrum J\"{u}lich and JARA, D-52425 J\"{u}lich, Germany}

\date{\today}
\vspace{1cm}

\begin{abstract}
We present a first-principles computational scheme for investigating the ballistic transport properties
of one-dimensional nanostructures with non-collinear magnetic order. The electronic structure is obtained
within density functional theory as implemented in the full-potential linearized augmented plane-wave (FLAPW)
method and mapped to a tight-binding like transport Hamiltonian via non-collinear Wannier functions. 
The conductance is
then computed based on the Landauer formula using the Green's function method.
As a first application we study the conductance between two ferromagnetic Co monowires terminated by single
Mn apex atoms as a function of Mn-Mn separation.
We vary the Mn-Mn separation from the contact (about $2.5$ to $5$~{\AA}) to the far tunneling regime ($5$ to $10$~{\AA}).
The magnetization
direction of the Co electrodes is chosen either in parallel or antiparallel alignment and we allow for
different spin configurations of the two Mn spins. In the tunneling and into the contact regime the
conductance is dominated by $s$-$d_{z^2}$-states. In the close contact regime (below $3.5$~{\AA}) there
is an additional contribution for a parallel magnetization alignment from the $d_{xz}$- and $d_{yz}$-states
which give rise to an increase of the magnetoresistance as it is absent for antiparallel magnetization.
If we allow the Mn spins to relax a non-collinear
spin state is formed close to contact due to the competition of ferromagnetic coupling between Mn and Co and
antiferromagnetic coupling between the Mn spins. We demonstrate that the transition from a collinear to such
a non-collinear spin structure as the two Mn atoms approach leaves a characteristic fingerprint in the distance-dependent
conductance and magnetoresistance of the junction. We explain the effect of the non-collinear spin state on the
conductance based on the spin-dependent hybridization between the $d_{xz,yz}$-states of the Mn spins and their
coupling to the Co electrodes.
\end{abstract}

\maketitle

\section{Introduction}
\label{sec:Introduction}

Break junction experiments have allowed to perform transport studies on nanoscale metallic contacts
in which the mean free path of the electrons is much larger than the junction length. The observation
of quantized conductance in such systems is a hallmark of ballistic transport and opened new vistas to
study the scaling of electronic devices down to the atomic length scale.\cite{Yanson} A drawback of such experiments is
the limited control of the microscopic arrangement in the junction which hinders a straight forward
interpretation of the data and makes a comparison with theoretical calculations difficult.\cite{ThiessNL2008} In this
respect, a great advantage is given by the use of scanning tunneling microscopy (STM) experiments, 
in which a tip can approach and contact single atoms or molecules on a surface.\cite{NeelPRL2009,KroegerJPCM2008,TaoPRB2010,ChopraNM2005,CalvoN2009,ZieglerNJP2011}
In such experiments, it has been possible to measure
the conductance as a function of tip-sample distance from the tunneling to the contact regime. Due to the
promise of spintronic devices for future applications with low power consumption and high speed, a recent
focus of such contact measurements has been magnetic systems, e.g.~spin-valve behavior has been observed
in single magnetic molecules or atoms on surfaces~\cite{ZieglerNJP2011,WulfhekelNN2011} and the occurrence
of the Kondo effect has been found in ferromagnetic atomic contacts.\cite{CalvoNature2009}

It has been emphasized that the low coordination of the contact atoms in nanoscale junctions leads
to an enhanced tendency towards magnetism, e.g.~magnetic moments are formed in systems of otherwise non-magnetic materials.\cite{PhysRevB.68.144434,Delin2004,SmogunovNN2008,Alex2009,Mokrousov2010} 
Naturally, transport phenomena in such magnetic low-dimensional systems have raised
a lot of attention and triggered many theoretical studies, which mainly focused on systems with
collinear magnetic order, considering also the effect of magnetoresistance.\cite{Smogunov:2004,Smogunov:2006,Bagrets:2004,Bagrets:2007,PolokPRB2011,transportpaper} 
It was also recently realized that, if the magnetization direction of the two electrodes is opposite, a domain wall can form 
in the contact between them and the non-collinear order in the domain  strongly affects the conductance and the magnetoresistance.\cite{BurtonPRB2006,CzernerPRB2008,CzernerPSS2010} Finally, the effect of
spin-orbit coupling on the conductance needs to be considered\cite{Smogunov:2008,transportpaper} 
which leads to novel transport phenomena such as
the ballistic anisotropic magnetoresistance\cite{VelevPRL2005,transportpaper} or the tunneling anisotropic
magnetoresistance.\cite{Bolotin2006,BurtonPRB2007}

Recently, the transition regime from tunneling to contact in a spin-polarized STM geometry has been studied based
on density functional theory in order to explain e.g.~the conductance of a single magnetic atom,\cite{TaoPRB2010}
and to analyze the contribution from different conduction channels.\cite{PolokPRB2011}
As a magnetic STM tip approaches a single magnetic atom on a surface an exchange interaction with the
tip apex atom occurs. In principle, it is possible to switch the magnetic moment of the adatom in such
a way.\cite{TaoPRL2009} If the magnetic moment of the adatom is exchange coupled to the substrate (as in
Ref.~\onlinecite{ZieglerNJP2011}) there is a competition of exchange interactions which can result in a
canting of the spins close to contact. Non-collinear spin alignment in such an atomic contact can also occur
if the adatom spin is canted due to exchange coupling on a substrate with a spin spiral structure as in
Refs.~\onlinecite{SerrateNN2010} and \onlinecite{Menzel:2012}. The effect of such a non-collinearity in the spin direction of the
tip apex and the adatom on the conductance
is the focus of the present work.

We introduce an approach to calculate the conductance in magnetic nanojunctions with non-collinear spin structure
from first-principles, employing the methodology of non-collinear Wannier functions (WFs), which we describe in detail.
 In order to start from an accurate description of the electronic and magnetic structure of
the system we use the full-potential linearized augmented plane wave (FLAPW) method based on density functional
theory. We map the electronic structure of a system in a non-collinear magnetic state from the FLAPW description
to a tight-binding like Hamiltonian via WFs. Finally, we calculate the conductance
within the Landauer approach with the technique of Green's functions.

As a model system, we consider two Co monowires to each of which a single apex Mn atom is attached. We vary
the distance between the two Mn atoms in order to calculate the conductance from the tunneling to the contact
regime. The magnetization direction of the two Co electrodes is chosen either parallel (P) or antiparallel (AP)
which allows us to obtain the distance-dependent magnetoresistance. In the tunneling regime the conductance is
dominated by states of $s$-$d_{z^2}$-orbital character and only in the contact regime there is an additional
contribution due to $d_{xz,yz}$-states. As the latter conduction channel is suppressed in the AP alignment the
magnetoresistance displays a large rise close to contact.

When the two Mn atoms approach in the P electrode alignment
a competition of the exchange interactions between the two Mn spins and the Mn spins with the Co electrodes occurs.
While the Mn spins couple ferromagnetically to the Co electrodes, they couple antiferromagnetically with each
other. As a result a non-collinear arrangement becomes the magnetic ground state and the angle between the two
Mn spins changes gradually from zero to about $105^\circ$ at the closest separation we considered. The conductance
displays a characteristic dip as the non-collinear state forms which is also apparent in the distance-dependent
magnetoresistance. We explain this reduction of the conductance due to non-collinear spin states from the
spin-dependent hybridization of $d_{xz,yz}$-states between the two Mn atoms which depends on the angle between
their spin moments and partly suppresses the conduction in this channel.

The paper is organized as follows. In Sec.~\ref{sec:Method} we introduce our method to calculate the conductance of
a one-dimensional nanoscale junction with a non-collinear spin structure. We discuss the extension of Wannier functions
to systems with non-collinear order (Sec.~\ref{sec:nocoWFs}), the implementation within the FLAPW method (Sec.~\ref{sec:FLAPW_to_WFs}),
and the incorporation into our transport code (Sec.~\ref{sec:nocoTrans}). In Sec.~\ref{sec:Co-Mn-Mn-Co} we introduce our
model system consisting of two Co monowires to each of which a single Mn atom is attached. First, we analyze the
magnetic and transport properties of collinear spin states from tunneling to contact (Sec.~\ref{subsec:t-to-c})
before we address the occurrence of non-collinear spin states in the contact regime (Sec.~\ref{subsec:origin_noco}).
We analyze the ballistic conductance of such spin states (Sec.~\ref{subsec:noco_junction}) and
show that a characteristic fingerprint is observed in the distance-dependent conductance and the magnetoresistance
(Sec.~\ref{subsec:t-to-c_rev}). We end with a summary in Sec.~\ref{sec:Summary}.

\section{Method}
\label{sec:Method}

The density functional theory (DFT)\cite{PhysRev.136.B864} states that the energy functional of a general magnetic system
$E[n(\mathbf{r}),\mathbf{m}(\mathbf{r})]$ is uniquely determined
by the charge density $n(\mathbf{r})$ and the magnetization density $\mathbf{m}(\mathbf{r})$.
The most common
approximation made to a general magnetic system is to
assume a collinear magnetization density, i.e.,
$\mathbf{m}(\mathbf{r})=m(\mathbf{r})\hat{\mathbf{e}}$,
where $\hat{\mathbf{e}}$ is an arbitrary direction.
Within this collinear approximation
the energy is a unique functional of the charge
density $n(\mathbf r)$ and the scalar magnetization density
$m(\mathbf r)$. Due to decoupled spin and real space, the spin-channels can be
treated independently. However, it is known that relaxing the collinear approximation
and allowing for non-collinearity of the magnetization density in real space in the DFT setup leads
to an ability of reliably treating whole classes of new phenomena, which rely on the properties of complex
magnetic states.\cite{KurznocoFLAPW}

\subsection{Non-collinear Wannier Functions}
\label{sec:nocoWFs}

Within the DFT formulation for non-collinear magnetic systems one solves the
Kohn-Sham equations\cite{KurznocoFLAPW}
\begin{equation}
\label{sKSEq}
\{-\frac{\hbar^2}{2m_e}\nabla^2\mathbf{I}_2+\mathbf{V}\}\Ket{\mbox{\boldmath{$\psi$}}_{\mathbf km}}=
\mbox{\boldmath{$\epsilon$}}_m(\mathbf k)\Ket{\mbox{\boldmath{$\psi$}}_{\mathbf km}},
\end{equation}
where $\mathbf{I}_2$ is the $2\times2$
unity matrix, $\mathbf{V}$ is the potential matrix which also mixes the spin channels, $m$ is the band index and
$\Ket{\mbox{\boldmath{$\psi$}}_{\mathbf k m}}=(\Ket{\psi_{\mathbf k m \uparrow}},\Ket{\psi_{\mathbf k m \downarrow}})^{T}$
is the spinor Bloch function
with spin-up and spin-down components $\Ket{\psi_{\mathbf k m \uparrow}}$ and $\Ket{\psi_{\mathbf k m \downarrow}}$, respectively.

For the $M$ converged spinor Kohn-Sham orbitals $\Ket{\mbox{\boldmath{$\psi$}}_{\mathbf km}}$
on a uniform mesh of $\mathcal{N}$ $\mathbf k$-points, the orthonormal set of Wannier
functions can be obtained via the transformation\cite{WannierPR1937}
\begin{equation}
\label{Umn}
\Ket{\mathbf W_{\mathbf Rn}}=\frac{1}{\mathcal{N}}\sum_{\mathbf k}e^{-i\mathbf
  k\cdot\mathbf R}\sum_{m=1}^MU_{mn}^{(\mathbf k)}\Ket{\mbox{\boldmath{$\psi$}}_{\mathbf km}},
\end{equation}
where the number of WFs $N$ is smaller than or equal to $M$ and
the matrices $U_{mn}^{(\mathbf k)}$ represent the gauge freedom of the WFs.
In case when $N=M$ and the group of bands we
are extracting the WFs from is isolated from other bands, the $U_{mn}^{(\mathbf k)}$-matrices
are unitary at each $k$-point. Imposing the constraint of maximal
localization of WFs in real space determines the set of $U_{mn}^{(\mathbf k)}$-matrices
up to a common global phase, and the corresponding WFs are
called maximally-localized Wannier functions (MLWFs).\cite{MarzariPRB1997}
Like the Kohn-Sham orbitals, the MLWFs are spinors and can be written as
$\Ket{\mbox{\boldmath{$W$}}_{\mathbf k m}}=(\Ket{W_{\mathbf k m \uparrow}},\Ket{W_{\mathbf k m \downarrow}})^{T}$
in terms of their spin-up and spin-down components $\Ket{W_{\mathbf k m \uparrow}}$ and $\Ket{W_{\mathbf k m \downarrow}}$, respectively.

For the construction of MLWFs within DFT electronic structure codes, the matrix elements
$M^{(\mathbf{k},\mathbf{b})}_{mn}=\Bra{
\psi_{\mathbf k m }}
e^{-i\mathbf b\cdot \hat{\mathbf r}}
\Ket{
\psi_{\mathbf k+\mathbf b n}}$
and
$A_{mn}^{(\mathbf{k})}
=\langle
\psi_{\mathbf k m }
|
g_n \rangle
$
need to be computed, where $\Ket{g_n}$
is a localized orbital, which defines the starting point of the iterative procedure of
determining the MLWFs.~\cite{MarzariPRB1997}
Since spin-up and spin-down are coupled in noncollinear
calculations, these matrix elements involve a summation over the spin $\sigma$.

The matrix elements $\mathbf H_{nn^\prime}(\mathbf R_1-\mathbf R_2)$ for the WFs tight-binding Hamiltonian
\begin{equation}
\label{H_WFs}
\hat{\mathbf H}_{\mathrm{WFs}}=\sum_{\mathbf{R}_1n}\sum_{\mathbf{R}_2n^\prime}
\mathbf H_{nn^\prime}(\mathbf R_1-\mathbf R_2)\Ket{\mathbf W_{\mathbf{R}_1n}}
\Bra{\mathbf W_{\mathbf{R}_2n^\prime}}
\end{equation}
are given by\cite{FreimuthPRB2008}
\begin{equation}
\label{noco_matrix_elements}
\begin{array}{ll}
\mathbf H_{nn^\prime}&(\mathbf R_1-\mathbf R_2)=\\&\\&\displaystyle{\frac{1}{\mathcal{N}}\sum_{\mathbf{k}m}}\mbox{\boldmath{$\epsilon$}}_m(\mathbf k)
\Braket{\mathbf W_{\mathbf{R}_1n}|\mbox{\boldmath{$\psi$}}_{\mathbf{k}m}}
\Braket{\mbox{\boldmath{$\psi$}}_{\mathbf{k}m}|\mathbf W_{\mathbf{R}_2n^\prime}}=
\\&\\&\displaystyle{\frac{1}{\mathcal{N}}\sum_{\mathbf{k}m}}\mbox{\boldmath{$\epsilon$}}_m(\mathbf k)
e^{i\mathbf k \cdot(\mathbf R_1-\mathbf R_2)}(U_{mn}^{(\mathbf{k})})^{*}(U_{mn'}^{(\mathbf{k})})
.
\end{array}
\end{equation}
Even though the Wannier and Bloch functions are spinor valued, the transformation of the Hamiltonian from Bloch into Wannier representation is
fully determined by the matrices $U_{mn}^{(\mathbf{k})}$ and the eigenvalues $\mbox{\boldmath{$\epsilon$}}_m(\mathbf k)$ as in the collinear case.

\subsection{Non-collinear Wannier Functions within the FLAPW method}
\label{sec:FLAPW_to_WFs}
The treatment of noncollinear magnetism within
the FLAPW method
as implemented in the J\"ulich DFT code \texttt{FLEUR}\cite{fleur,KurznocoFLAPW}
neglects the effect of intra-atomic non-collinearity. Space is partitioned into
the muffin-tin (MT) and interstitial regions (IR).
The spin density $\mathbf{m}(\mathrm{r})$ in the IR is treated without shape
approximation as a continuous vector field. In the MT-sphere $\mathrm{MT}^\alpha$ of atom $\alpha$ only the projection
of the spin density onto the direction $\mathbf{\hat{e}}^\alpha_M$ of the average spin moment is used for the generation of
the exchange-correlation potential. The explicit
one- and two-dimensional implementations also contain a third region, the vacuum region (VR), which
can be treated analogously to the IR.\cite{FilmFLAPW,Mokrousov:2005} Thus, the self-consistent spin density is approximated as
\begin{equation}
\label{m_FLAPW}
\mathbf{m}(\mathbf{r})=\left\{\begin{array}{ll}\mathbf{m}(\mathbf{r})&\mbox{IR (VR)}\\
m^\alpha(\mathbf{r})\mathbf{\hat{e}}^\alpha_M&\mbox{$\mathrm{MT}^\alpha$ }\end{array}\right..
\end{equation}
A part of the intra-atomic non-collinearity can still be described within this hybrid approach by decreasing
the MT radii.

The radial solutions $u^\alpha_{l\sigma^\alpha}(r)$ of angular momentum $l$ of the scalar-relativistic Schr\"{o}dinger equation
in $\mathrm{MT}^\alpha$
and their energy derivatives $\dot u^\alpha_{l\sigma^\alpha}(r)$ are calculated for the two spins $\sigma^\alpha$ and used for
the expansion of basis functions and Bloch functions. The spin quantum number $\sigma^\alpha$ refers to the local
spin quantization axis $\mathbf{\hat{e}}^\alpha_M$. The expansion coefficients of the eigenspinors of the local spin quantization axis in
terms of the eigenspinors of the global spin quantization axis, which is the $z$ axis, are given by
\begin{equation}
\begin{array}{l}
\chi^{\alpha g}_\uparrow=(\exp{(-i\frac{\phi}{2})}\cos{(\frac{\theta}{2})},\exp{(i\frac{\phi}{2})}\sin{(\frac{\theta}{2})})^T\\
\chi^{\alpha g}_\downarrow=(-\exp{(-i\frac{\phi}{2})}\sin{(\frac{\theta}{2})},\exp{(i\frac{\phi}{2})}\cos{(\frac{\theta}{2})})^T,
\end{array}
\end{equation}
where $\phi$ and $\theta$ are azimuthal and polar angles of the spin direction of $\mathrm{MT}^\alpha$ with respect to the global frame $g$.

Within the MTs the wave function $\mbox{\boldmath{$\psi$}}_{\mathbf km}(\mathbf r)$ is thus given by:
\begin{equation}
\label{basis_FLAPW}
\begin{array}{ll}
\mbox{\boldmath{$\psi$}}_{\mathbf km}(\mathbf r)|_{\mathrm{MT}^\alpha}=\displaystyle{\sum_{\sigma^\alpha L}}&\left[A_{mL\sigma^\alpha}^{\alpha}(\mathbf k)
u^\alpha_{l\sigma^\alpha}(r)\right.\\&\\&\left.+B_{mL\sigma^\alpha}^{\alpha}(\mathbf k)\dot u^\alpha_{l\sigma^\alpha}(r))\right]
Y_{L}(\hat{\mathbf r}))\chi^{\alpha g}_{\sigma^\alpha}
\end{array}
\end{equation}
where $L$ denotes the angular momentum quantum numbers and $m$ is the band index.

Using functions $\Ket{g_n}$ which are restricted each to a single MT-sphere has been found to result in a very good starting point for the
iterative optimization of collinear WFs.~\cite{FreimuthPRB2008}
Due to the approximate intra-atomic collinearity, it is reasonable to choose in the noncollinear case the
localized orbitals $\Ket{g_n}$ to
be eigenstates of the projection of the spin-operator onto
the local spin-quantization axis $\mathbf{\hat{e}}^\alpha_M$:
\begin{equation}
\label{Amn}
g_{n}(\mathbf r)|_{\mathrm{MT}^{\alpha(n)} }
=\displaystyle{\sum_{L}}c_{nL} \tilde{u}_{nl}(r)  Y_{L}(\hat{\mathbf r}))\chi^{\alpha(n) g}_{\sigma(n)}.
\end{equation}
Here, $c_{nL}$ are expansion coefficients, $\alpha(n)$ is the index of the atom for which $g_{n}(r)$ is non-zero, $\sigma (n)$ is the
spin associated with this trial orbital $g_{n}(\mathbf{r})$ and $\tilde{u}_{nl}(r)$ is the radial part of the trial orbital.
Thus we obtain:
\begin{equation}
\begin{array}{ll}
A_{mn}^{(\mathbf{k})}=&\displaystyle{\sum_{L}} c_{nL}\\&\\
&\times\{[A^\alpha_{mL\sigma(n)}(\mathbf{k})]^*\int u_{l\sigma(n)}^{\alpha(n)}(r)\tilde{u}_{nl}(r)r^2\,dr\\&\\
&+[B^\alpha_{mL\sigma(n)}(\mathbf{k})]^*\int
\dot u_{l\sigma(n)}^{\alpha (n)}(r)\tilde{u}_{nl}(r)r^2\,dr\}.
\end{array}
\end{equation}

The MT contribution to the $M^{(\mathbf{k},\mathbf{b})}_{mn}$ matrix may be written as
\begin{equation}
M^{(\mathbf{k},\mathbf{b})}_{mn}|_{\mathrm{MT}}
=\sum_{\alpha,\sigma^{\alpha}}
\int_{\mathrm{MT}^\alpha}
d^3\,r
[\psi_{\mathbf{k}m\sigma^{\alpha}} (\mathbf{r}) ]^{*}
e^{-i\mathbf{b}\cdot\mathbf{r}}
\psi_{\mathbf{k}+\mathbf{b}n\sigma^{\alpha}}(\mathbf{r}),
\end{equation}
with $\psi_{\mathbf{k}m\sigma^{\alpha}}(\mathbf{r})$ given by
\begin{equation}
\begin{aligned}
&\psi_{\mathbf{k}m\sigma^{\alpha}}(\mathbf{r})=\sum_{L}Y_{L}(\hat{\mathbf r})\times\\
&\times\left[A_{mL\sigma^\alpha}^{\alpha}(\mathbf k)
u^\alpha_{l\sigma^\alpha}(r)+B_{mL\sigma^\alpha}^{\alpha}(\mathbf k)\dot u^\alpha_{l\sigma^\alpha}(r))\right]
.
\end{aligned}
\end{equation}
The computation of the $M^{(\mathbf{k},\mathbf{b})}_{mn}$ matrix for noncollinear systems reduces therefore to integrals for which
explicit expressions have been given for the FLAPW method.~\cite{FreimuthPRB2008}

\subsection{Ballistic transport in systems with non-collinear magnetism}
\label{sec:nocoTrans}

The extension of the collinear scheme for ballistic transport, which we described
in detail in our previous publication, Ref.~\onlinecite{transportpaper}, to a non-collinear setup is now rather straightforward.
Given the minimal WFs
Hamiltonian $\hat{\mathbf H}_{\rm{WFs}}$ (Eq.~(\ref{H_WFs})) for a non-collinear system, we are able to construct the
tight-binding Hamiltonian of the nanojunction in accordance
to our transport method,\cite{transportpaper} which employs the partitioning of space into the scattering region
$(S)$, as well as left $(L)$ and right $(R)$ leads:
\begin{equation}
\label{OpenSys}
\mathbf H=\left(
\begin{array}{ccc}
\mathbf H_{L}&\mathbf H_{LS}^\dagger&\mathbf 0\\
\mathbf H_{LS}&\mathbf H_{S}&\mathbf H_{SR}\\
\mathbf 0&\mathbf H_{SR}^\dagger&\mathbf H_{R}
\end{array}\right).
\end{equation}
Compared to a collinear calculation we have to deal with twice as many Wannier functions due to the inseparable
spin-channels.
Two calculations using the locking technique\cite{transportpaper} are required for Eq.~(\ref{OpenSys}). The Hamiltonian matrix $\mathbf H_S$ and the
matrices $\mathbf H_{LS}$, describing the coupling to the leads, are obtained from a super-cell FLAPW calculation.
The Green's functions
for the leads $\mathbf G_{L/R}(E)=[(E+i\epsilon)\mathbf I_{L/R}-\mathbf H_{L/R}]^{-1}$ can be brought to finite sized
surface Green's functions $\mathbf g_{L/R}$ by constructing $\mathbf H_{L/R}$ based on principal layers $\mathbf h_{L/R}$ and the coupling matrices
$\mathbf h_{LL/RR}$.\cite{transportpaper} Those matrices are obtained from a separate calculation of a perfect periodic lead.
Following our Landauer-B\"uttiker method the non-collinear ballistic transport can be calculated with the
Green's function of the scattering region:
\begin{equation}
\label{scatteringGF}
\mathbf G_{S}(E)=[E\mathbf I_S-\mathbf H_{S}-\mathbf H_{LS}^\dagger \mathbf g_{L}\mathbf H_{LS}-\mathbf H_{SR}^\dagger \mathbf g_{R} \mathbf H_{SR}]^{-1}.
\end{equation}
The interaction between scattering region and the leads, and the resulting level broadenings are described by broadening matrices $\Gamma$
\begin{equation}
\mathbf\Gamma_{L/R}(E)=i[\mathbf\Sigma_{L/R}(E)-\mathbf\Sigma_{L/R}^\dagger(E)],
\end{equation}
where $\mathbf\Sigma_{L/R}(E)$ are the self-energies of the leads:
\begin{equation}
\mathbf \Sigma_{L/R}(E)=\mathbf H_{LS/SR}^\dagger \mathbf g_{L/R}(E)\mathbf H_{LS/SR}.
\end{equation}
Finally the ballistic transport process is described by the transmission function $T(E)$
\begin{equation}
\label{Transmission}
T(E)={\rm Tr}[\mathbf G_S(E)\mathbf\Gamma_L(E)\mathbf G_S^\dagger(E)\mathbf\Gamma_R(E)].
\end{equation}
resulting in the conductance through the junction
\begin{equation}
G(E)=\frac{e^2}{h}T(E)=\frac{1}{2}G_0T(E)
\end{equation}
with the conductance quantum $G_0=2e^2/h$.
 In the non-collinear case the trace
operation of Eq.~(\ref{Transmission}) has to be additionally performed over the spin $\sigma$.
The spin-channel information is therefore lost for general non-collinear systems.

We tested non-collinear Wannier functions on freestanding non-collinear magnetic Mn chains and found them to reproduce the
FLAPW electronic structure with any given accuracy.
For the performed transport calculations a $2^{\mathrm{nd}}$ 
nearest-neighbor (NN) 
tight-binding like Hamiltonian is sufficient
due to the excellent correspondence of FLAPW and WFs electronic structure in the vicinity of the Fermi level in
absence of $s$-$d_{z^2}$ band edges in that particular region, which usually would require to consider
more neighbors \cite{transportpaper}. For the system, considered in the following, the orbitals participating in
transport can be arranged according to the symmetry into the $\Delta_1$ ($s$ and $d_{{z^2}}$ orbitals), the
$\Delta_3$ ($d_{{xz/yz}}$) and the $\Delta_4$ ($d_{{xy/x^2-y^2}}$) groups.

\section{The \cmmc junction}
\label{sec:Co-Mn-Mn-Co}

\begin{figure}
\begin{center}
\centerline{\includegraphics[width=0.45\textwidth,angle=0]{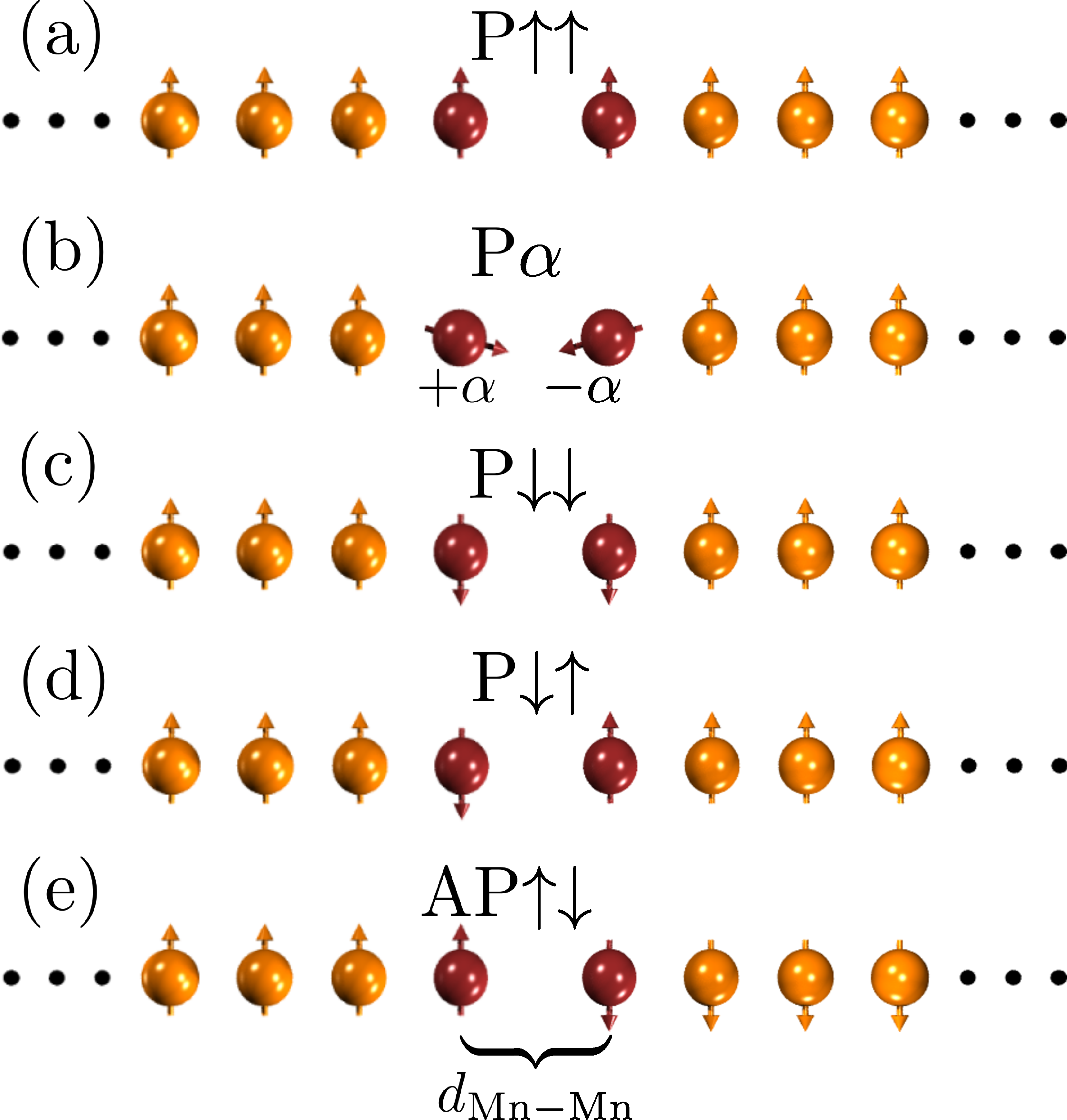}}
\caption{(color online) Calculated magnetic configurations in the \cmmc
 junction. The magnetization of the left lead always points up ($\uparrow$). The
 magnetization of the right lead can be either parallel (P) or anti-parallel (AP) to it.
 The direction of the Mn spins is marked with respect to the left lead: 
 pointing up
 ($\uparrow$), down ($\downarrow$), or along a direction at an angle $\alpha$ with
 the magnetization of the left lead. 
 Special configurations of interest which are 
 discussed in the text are shown: (a) P\up\up, (b) symmetric P$\alpha$ state with
 $\alpha=105^{\circ}$, P$105^\circ$, (c) P\down\down~(d)
P\down\up~and (e) AP\up\down. In the text, we refer to the separation between the
leads in terms of the distance between the Mn atoms, $d_{\rm{Mn-Mn}}$. Co atoms 
are displayed as orange spheres, Mn
atoms are displayed as red spheres.
}
\label{fig:config}
\end{center}
\vspace*{-1cm}
\end{figure}

In the following section, we investigate the ballistic transport properties of collinear and
non-collinear magnetic configurations of a \cmmc  junction, consisting of semi-infinite ferromagnetic Co monowires with
magnetic Mn "tip" atoms, see sketch of the structure in Fig.~\ref{fig:config}.
We will discuss the effect of non-collinear
magnetism on ballistic transport through such a junction specifically keeping in mind tunneling-to-contact
STM\cite{NeelPRL2009,KroegerJPCM2008,TaoPRB2010,ChopraNM2005,CalvoN2009,ZieglerNJP2011}
and mechanically-controllable break-junctions\cite{KizukaPRB2008} experiments. In particular, we investigate the changes in the transport properties upon changing the
distance between the two Mn atoms, while keeping all other interatomic distances
fixed at their equilibrium "semi-infinite" values. We will show, that upon bringing the
leads together the non-collinearity in this system emerges as a result of competing
Mn-Mn and Mn-Co exchange interactions.

We then explore the influence of non-collinear magnetism on ballistic transport for various collinear and
non-collinear configurations.
The nomenclature for the magnetic states in the \cmmc junction includes the
alignment of the magnetization directions of the leads, parallel (P) or anti-parallel (AP)
to each other, and the directions of the two Mn spins. Without loss
of generality, these directions are denoted with respect to the left lead which
has magnetization "up". The Mn spins can point "up" ($\up$), "down" ($\down$) or
in a direction which makes an angle $\alpha$ with the direction "up",
see Fig.~\ref{fig:config}. In the latter case we consider the symmetric configuration,
denoted as P$\alpha$,
in which the spins of the Mn atoms make an angle of $2\alpha$ between each other.
For all considered non-collinear P$\alpha$ states we fixed the direction of all Co atoms
either up or down, depending on the magnetization direction of the corresponding
lead. The energy differences between different magnetic states are given per Mn atom.

\subsection{Computational details}
\label{subsec:CD}

For all collinear and non-collinear electronic structure calculations we used density functional theory within generalized gradient approximation (GGA) to the exchange-correlation potential,\cite{rpbe} as implemented in the FLAPW J\"ulich code \texttt{FLEUR}.\cite{fleur}
The wires were calculated in three-dimensional super-cells, with an interchain separation in the $x$-$y$$-$plane of 13~bohr. The super-cell setup along the chain's axes ($z$-direction) is described in detail below.
The Brillouin zone (BZ) was sampled by 12 or 24 $k$-points along the $z$-axis, depending on the size of the super-cell.
All calculations were performed with an LAPW basis cut-off parameter $k_{\mathrm{max}}$ of 3.7 bohr$^{-1}$,
resulting in approximately 625 LAPW basis functions per atom.

The parallel magnetic configuration (P) of \cmmc junctions was investigated in an 8-atom super-cell along the
chain direction, consisting of six Co atoms with an equilibrium interatomic distance of the Co infinite monowire
of $d_{\rm{Co}}=4.15$~bohr, and two attached Mn atoms, see Fig.~\ref{fig:config}. For all considered magnetic
configurations, with parallel or
antiparallel alignment of the magnetization of the leads, as well as non-collinear magnetic states, irrespective of
the separation between the leads, we fixed the Co-Mn distance $d_{\rm{Co-Mn}}$ to 4.48~bohr, which
corresponds to the equilibrium distance between the ferromagnetic Co and Mn atoms at a very large separation
 between the leads.  For the P\up\up~and P\down\down-states of the junction we considered the inter-Mn separation of $d_{\rm{Mn-Mn}}=5.0$, $5.5$, $7.0$, $8.5$, $10.0$, $12.5$, $15.0$, $17.5$ and $20.0$~bohr.
The anti-parallel magnetic configuration (AP) of \cmmc junctions was calculated in a 16-atom super-cell, consisting of six $\uparrow$-Co atoms and two Mn atoms on one side, and six $\downarrow$-Co atoms
and two Mn atoms at another end of the junction. In this case $d_{\rm{Mn-Mn}}$ was set to 4.5, 5.0, 5.5,
7.0 and 8.5~bohr.

For the conductance calculations we applied the locking technique to a
perfect monowire to describe the semi-infinite leads, as described in detail in Ref.~\onlinecite{transportpaper}.
In all cases the Wannier functions were generated on a $1\times 1\times 24$ $k$-point grid in the BZ.
For the collinear cases the WFs were generated from one $4s$- and five $3d$-orbitals per atom for each
spin separately, which were constructed from the radial solutions for the FLAPW potential.
In non-collinear 
calculations 
the spin channels are mixed, and two $4s$- and ten
$3d$-orbitals per atom were used to construct the WFs per atom. The energy bands were disentangled
using the procedure described in Ref.~\onlinecite{disentanglement}. For the collinear
calculations the lowest 58 eigenvalues per $k$-point were used
to obtain 48 WFs for the 8 atom super-cell and the lowest 104 eigenvalues per $k$-point for
96 WFs for the 16 atom super-cell calculations. With non-collinearity of the magnetization included
the lowest $103$ eigenvalues per $k$-point were used to obtain 96 WFs for the 8 atom unit cell.
For testing purposes, for several non-collinear configurations we compared the electronic
structure of the system calculated with \texttt{FLEUR} and with corresponding WFs, finding that a very
good description of the electronic structure can be achieved with WFs  within the 3rd nearest-neighbor
approximation, while for WFs calculations of the transmission in the vicinity of the Fermi level already the
2nd nearest-neighbor approximation to the WFs Hamiltonian provides very reliable results.

\subsection{Collinear magnetic states of the junction from tunneling to contact}
\label{subsec:t-to-c}
\begin{figure}
\begin{center}
\centerline{\includegraphics[width=0.48\textwidth]{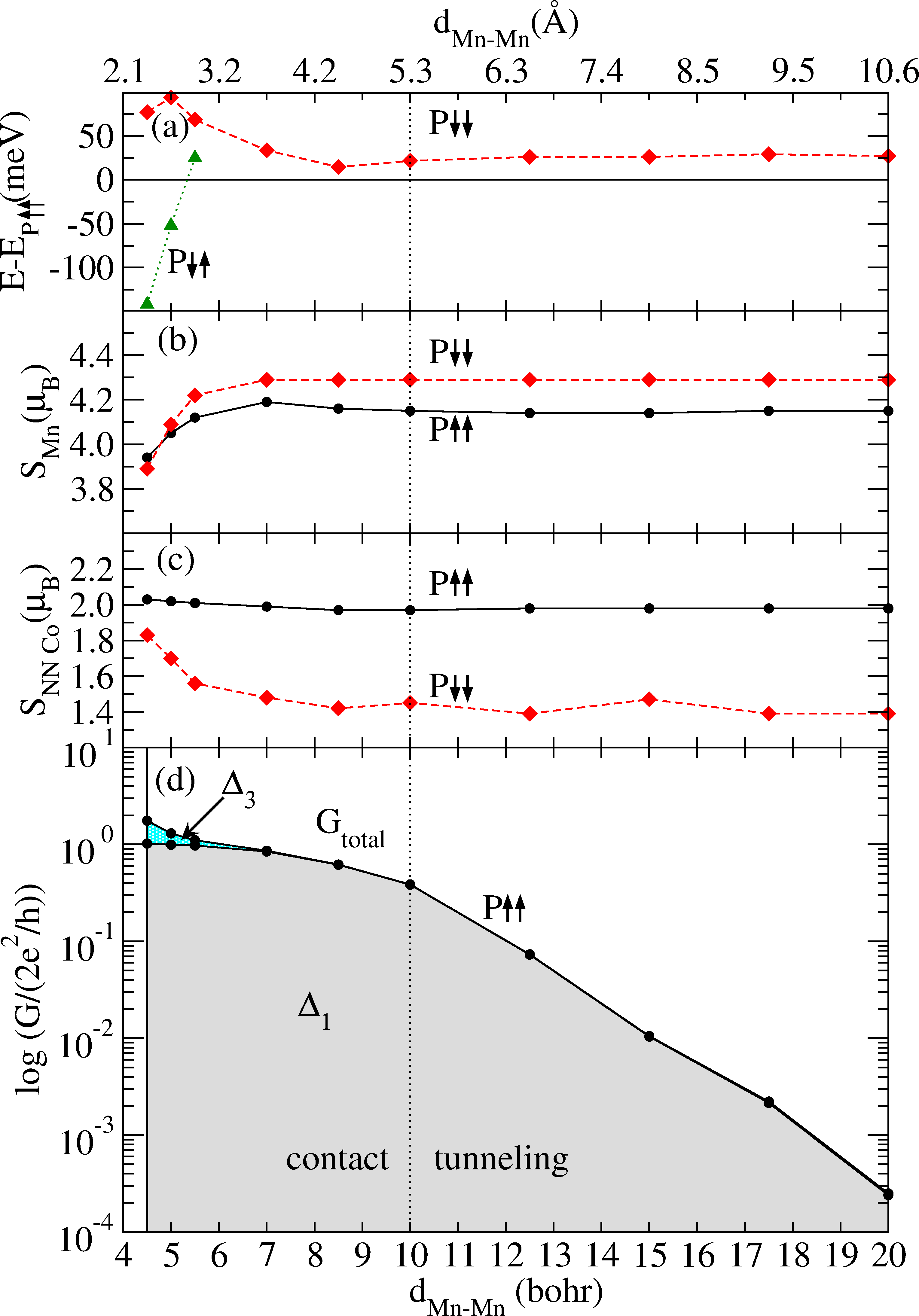}}
\caption{(color online) (a) Energy of the P\down\down~state (red diamonds, dashed line) and  P\down\up~state (green triangles, dotted line) with respect to the energy of the P\up\up~state, as a function of the Mn-Mn interatomic distance
$d_{\rm{Mn-Mn}}$. Spin moment of the Mn atoms, \SMN, (b), and NN Co atoms, \SCO, (c), for P\up\up~(black circles, solid line) and P\down\down~(red diamonds, dashed line) configurations are given as a function of $d_{\rm{Mn-Mn}}$.
(d) Total conductance at the Fermi level, $G_{\rm{total}}(E_F)$ (upper line) and $\Delta_1$ conductance
$G_{\Delta_1}(E_F)$ (lower line) on a logarithmic scale for the P\up\up-state. Gray shaded area is associated with the
$\Delta_1$ conductance, while cyan shaded area with the $\Delta_3$ conductance.
}
\label{fig:stretch}
\end{center}
\end{figure}

We start the investigation of the \cmmc junction with both leads positioned far away from each other.
To mimic a tip-sample approach, we decrease the Mn-Mn distance $d_{\rm{Mn-Mn}}$, and calculate the energies of 
the collinear states P$\up\up$, P$\down\down$ and P$\down\up$, showing the results in Fig.~\ref{fig:stretch}(a).
The energy difference between P\up\up~and P\down\down~states when the distance is varied 
in the tunneling regime
from $d_{\rm{Mn-Mn}}=20$~bohr down to $10$~bohr remains relatively constant and constitutes around $27$~meV per atom, indicating
weak interaction between both sides of the junction and a weak ferromagnetic coupling between the Mn atom
and its nearest Co neighbor (NN Co).
After a small reduction of the energy difference between the P\up\up~and P\down\down~states
around $d_{\rm Mn-Mn}=8$~bohr,
the ferromagnetic (FM) Mn-Co coupling becomes more stable for decreasing $d_{\rm{Mn-Mn}}$, expressed in an increasing energy difference. In the contact regime we find
a slight decrease in the energy difference from 93.5~meV at $d_{\rm{Mn-Mn}}=5.0$~bohr down to $77$~meV per
atom at $d_{\rm{Mn-Mn}}=4.5$~bohr. This decrease in energy can be correlated with strong changes
in the Mn and NN~Co spin moments, $S_{\rm{Mn}}$ and $S_{\rm{Mn}}$, respectively,  upon decreasing the
distance, see Fig.~\ref{fig:stretch}(b) and (c) (see also discussion in the next section).

While in all cases the spin moments of the Co atoms, not neighboring the Mn atoms directly ($\approx 2.09\,\mu_B$) are very similar to the spin moments of the Co atom in an infinite lead ($\approx 2.07\,\mu_B$), the spin moments of Mn atoms and the NN Co atoms  can be strongly affected by $d_{\rm{Mn-Mn}}$ 
at close contact and the spin configuration 
of the junction.
Namely, for the P\down\down~state $S_{\rm{Mn}}$ decreases from 4.3~$\mu_B$ to
3.9~$\mu_B$, while~NN~$S_{\rm{Co}}$ increases from $1.4\,\mu_B$ to $1.8\,\mu_B$, as $d_{\rm{Mn-Mn}}$ is varied
from 5.5 to 4.5~bohr. On the other hand, if the Mn spin moment exhibits a similar variation as a function of distance for the P\up\up~state,
the spin moment of the~NN~Co atoms remains relatively constant ($\approx 2\,\mu_B$).
This interplay between structure and magnetism already indicates that the intra-atomic as well
as inter-atomic exchange, given by the Stoner parameter $I$ and the Heisenberg exchange constants $J$, respectively, may be of importance for further understanding of the magnetic properties of this system.

The change from FM coupling at larger interatomic distances to an antiferromagnetic (AFM) coupling at smaller
$d_{\rm{Mn-Mn}}$ in an infinite Mn chain has been 
previously predicted based on DFT calculations.\cite{PhysRevB.75.104413} In the vicinity of this
crossover point the Mn spins favor non-collinear magnetic order.\cite{franziska,zeleny,ataca}
To demonstrate a strong
tendency of Mn spin moments to AFM coupling at smaller values of $d_{\rm{Mn-Mn}}$ we
plot the energy difference between the P\up\up~and P\down\up~states in
Fig.~\ref{fig:stretch}(a). Reversing one of the Mn spin moments in the P\up\up~configuration is clearly
energetically more favorable than the P\up\up~state when the distance between the Mn atoms is below
$\approx 5.2$~bohr. In this case the gain in energy due to switch of the Mn spin moment can be explained only by
the strong AFM coupling of the two Mn atoms for this regime of interatomic distance, since the coupling
of the Mn atom with its~NN~Co atom is ferromagnetic.
The spin moments in the P\down\up~state at $d_{\rm{Mn-Mn}}=4.5$~bohr constitute $4.1\,\mu_B$ for Mn and $1.5\,\mu_B$ for its~NN~Co on the AFM side, and $3.7\,\mu_B$ for Mn and $1.9\,\mu_B$ for the~n.n~Co on the FM side.
For larger $d_{\rm{Mn-Mn}}$ values the P\up\up~state
is the lowest in energy 
as compared to all possible collinear states of the junction in which the magnetization
direction of the left and right leads is the same, which is indicative of the FM Mn-Mn coupling for larger
distances.

In Fig.~\ref{fig:stretch}(d) we present the results of our calculations for
the evolution of the ballistic conductance of a P\up\up~junction when going
from the tunneling to the contact regime. 
The main contribution to the conductance
at large Mn-Mn distances is coming solely from the $\Delta_1$ channel, owing
to the overlap between the $s$-$d_{z^2}$ orbitals of the neighboring Mn atoms
across the barrier. Within our approach, the expected exponential behavior of the conductance at 
very large distances is very nicely reproduced. 
At a distance of $d_{\rm{Mn-Mn}}\approx 10$~bohr the conductance approaches the magnitude 
of the conductance quantum, reaching saturation upon further decreasing the distance. 
For distances in the contact regime below 7~bohr 
more localized $d$-orbitals of $\Delta_3$  symmetry start contributing to the total conductance,
as can be seen in Fig.~\ref{fig:stretch}(d). The $\Delta_3$ contribution to the conductance
increases with decreasing distance. As we shall see in the following the details of
hybridization between the $\Delta_3$ orbitals are very sensitive to the magnetic
state of the junction. On the other hand, in all considered cases the $d$-states of $\Delta_4$ symmetry do not contribute to the conductance due to an
energetic mismatch between the states of this symmetry of NN Co and Mn atoms, see discussion
in section~\ref{subsec:noco_junction}.

\subsection{Non-collinear magnetic states of the junction in contact regime}
\label{subsec:origin_noco}

According to the findings presented above, we expect that Mn spin moments in the contact regime will experience a frustration when the magnetizations of the leads are parallel to each other. In this case, when Mn atoms are close enough, FM coupling of Mn spins
with NN Co atoms and AFM Mn-Mn coupling can possibly lead to a stable non-collinear magnetic state.
In order to consider this situation, we introduce an angle $\alpha$ between the spin moments of the Mn and the NN Co atoms, rotating the
first Mn spin moment by $+\alpha$ and the second one by $-\alpha$, while keeping the moments of the
Co atoms fixed, see Fig.~\ref{fig:config}(b). This is what we call a symmetric  P$\alpha$-state. We choose a
distance of 4.5~bohr between the Mn atoms as a representative
of the contact regime at which the Mn-Mn coupling is strongly antiferromagnetic.

\begin{figure}
\begin{center}
\centerline{\includegraphics[width=0.47\textwidth,angle=0]{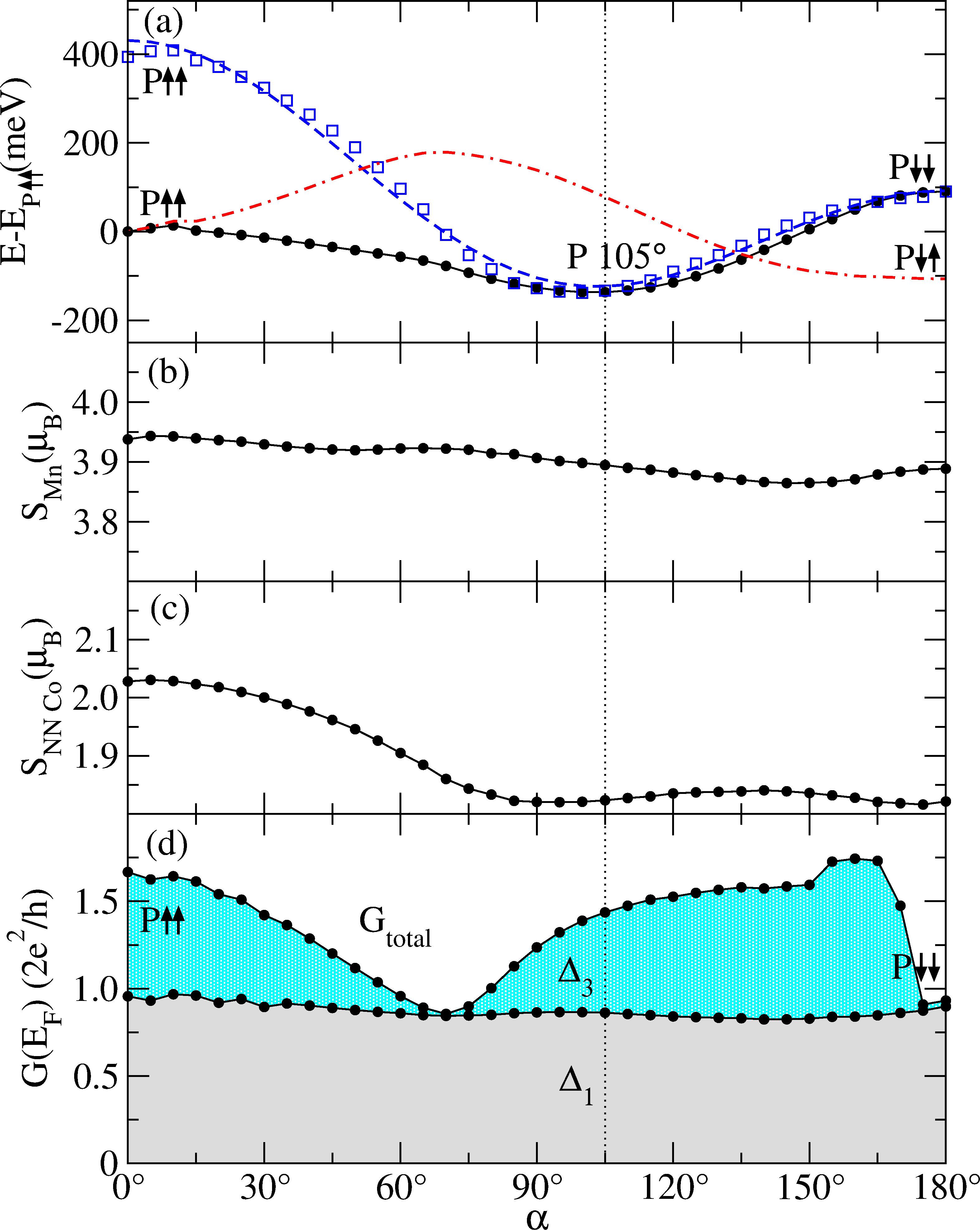}}
\caption{(color online) (a)
Total energy of the P$\alpha$ state 
with respect to the P\up\up~state as a function of angle $\alpha$ (solid line, filled circles). With 
open
black squares we show the total energy of the P$\alpha$ state from which the Stoner energy due to creation of the NN Co spin moments is subtracted. The corresponding fit of the corrected energy to the
Heisenberg model is shown with 
a dashed blue line. With the red
dotted line we show the total energy of the P\up\down~state with
only 
the 
$\down$-Mn spin rotated by an angle $\alpha$, calcualted within the Heisenberg model and incorporating also
the Stoner exchange.
(b) Mn spin moment as a function of angle $\alpha$. (c) NN Co spin moment as a function of angle $\alpha$.
(d) 
Ballistic conductance at the Fermi level 
for the P$\alpha$ state 
as a function of angle $\alpha$: total conductance $G_{\rm{total}}(E_F)$ (upper line), the $\Delta_3$ conductance contribution (cyan shaded area) and and $\Delta_1$ conductance $G_{\Delta_1}(E_F)$ (lower line, gray shaded area).
}
\label{fig:rotate}
\end{center}
\end{figure}

The results of our calculations for the total energy of the P$\alpha$ state, $E_{{\rm P}\alpha}$, in relation to the energy of the
P\up\up~state are shown in Fig.~\ref{fig:rotate}(a) as a function of the angle $\alpha$. From this plot we observe that the minimum of the total energy is acquired for the non-collinear P$105^{\circ}$ state, which is $137$~meV lower in energy than 
the 
corresponding collinear P\up\up~state. The failure of 
a straightforward description of the energy landscape $E_{{\rm P}\alpha}$ 
in terms
of a simple Heisenberg model which assumes just the nearest-neighbor Co-Mn and Mn-Mn exchange coupling,
given by antiferromagnetic $J_{\rm{Mn}}<0$ and ferromagnetic $J_{\rm{Co}}>0$, respectively, can be understood from noticing that the expression for
the energy within this approximation, given by
\begin{equation}
\label{Heisenberg}
E_{{\rm P}\alpha}(\alpha)= -\frac{1}{2}(J_{\rm{Mn}}\cos (2\alpha)+2J_{\rm{Co}}\cos (\alpha)),
\end{equation}
acquires a minimum for angles $\alpha$ below $90^{\circ}$, in contradiction to our calculations.

The solution to this 
deficiency of the Heisenberg model
can be given by lifting the assumption that the exchange interaction
between the Mn and Co spins, given by $J_{\rm Co}$, is ferromagnetic. As we can see from Fig.~\ref{fig:rotate}(b) and (c),
while the Mn spin moment remains relatively constant upon changing $\alpha$, \SCO~for values of $\alpha$ below
60$^{\circ}$ is by as much as $0.2\,\mu_B$ larger than for $\alpha>90^{\circ}$. Owing to the intra-atomic Stoner exchange, the non-collinear states with small $\alpha$ therefore acquire a negative contribution to the total energy in addition to that proportional
 to $J_{\rm Co}$, as compared to larger angles.
If we account for the energy gain due to creation of the NN Co spin moments by a Stoner parameter of Co, $I\approx 990$~meV, \cite{Gunnarson_ICo}
and subtract the energy gain $E_{\rm{Stoner}}=\frac{1}{2}I\SCO^2$ from the calculated DFT dispersion, we arrive at
the energy dispersion (squares in Fig.~\ref{fig:rotate}(a)), which reflects only exchange interactions between
the atoms. If we fit 
this curve according to 
Eq.(\ref{Heisenberg})
(dashed line in Fig.~\ref{fig:rotate}(a)), we obtain
the "non-renormalized" Heisenberg exchange constants of $J_{\rm Co}=-170$~meV and $J_{\rm{Mn}}=-366$~meV.
It becomes clear now, that, although the "pure" exchange coupling between the Mn and Co spins is expectedly
antiferromagnetic, the larger spin moment of Co 
in the parallel spin alignment with Mn,
tips the balance in favor
of ferromagnetic coupling between the spins, which can be observed for a large range of distances
$d_{\rm{Mn-Mn}}$,~c.f.~Fig.~\ref{fig:stretch}(a).

In Fig.~\ref{fig:stretch}(a) we also observe that, judging from the energies, in the close contact regime the collinear
P\down\up-state is competing with the non-collinear P$\alpha$-state for the global ground state of the system.
Indeed, our calculations show that at the  $d_{\rm{Mn-Mn}}$ of 4.5~bohr  the P\down\up-configuration is by
a tiny value of 5~meV 
lower in energy than the P105$^{\circ}$ solution. We argue, however, that the P\down\up-state is
not very likely to appear in experiments, given that the Co electrodes are identical. In this case, the adiabatic
rise of the intrinsically asymmetric P\down\up-configuration via symmetric non-collinear states cannot happen,
as the electrodes, initially being in the P\up\up-state when very far from each other, are brought together (see
also discussion at the beginning of section~\ref{subsec:t-to-c_rev}). Nevertheless, it seems plausible, that such
state, if observed in experiment, is created via a rapid flip of one of the Mn atoms in the contact regime, during,~e.g.,
a reformation of the lead geometry, or an inelastic current-induced spin-flip process. Our calculations, shown with a
dotted line in Fig.~\ref{fig:rotate}(a), based on
the Heisenberg model
extended by the Stoner term of intra-atomic exchange of the Co moments, 
indicate, that once the system enters the P\down\up-state, it is effectively "trapped" there, since the \down-Mn is
energetically quite stable versus deviations in the angle its spin makes with the rest of the spins in the system. Thus,
we do not consider any non-collinear states associated with the P\down\up-state in the following.

\subsection{Ballistic conductance of non-collinear magnetic states of the junction}
\label{subsec:noco_junction}

\begin{figure*}
\begin{center}
\centerline{\includegraphics[width=0.89\textwidth,angle=0]{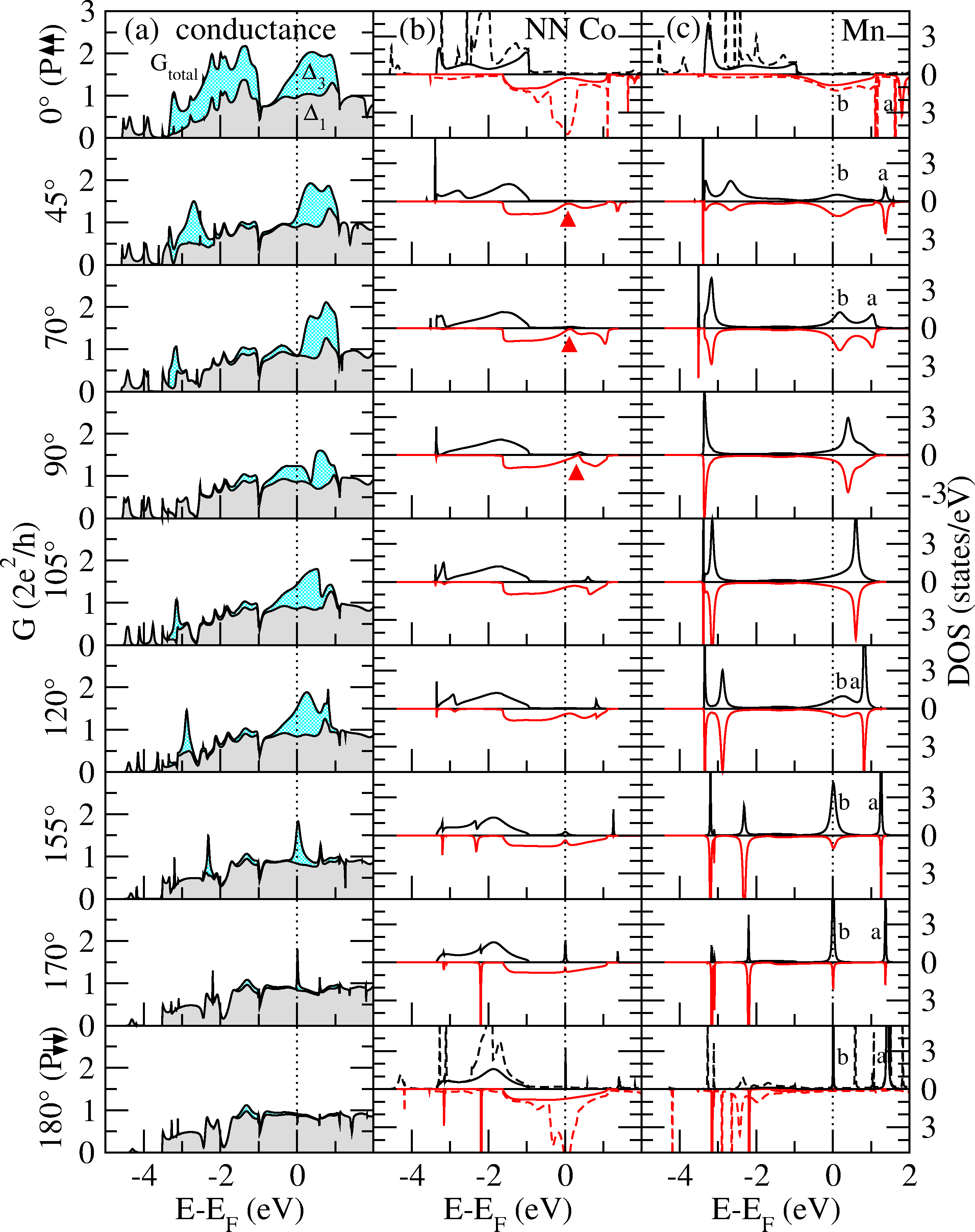}}
\caption{(color online) Transport properties and electronic structure of the P$\alpha$-state at $d_{\rm{Mn-Mn}}$ of 4.5~bohr
as a function of angle $\alpha$ (indicated on the left).
(a) Total conductance $G_{\rm{total}}$ (upper line) decomposed into the $\Delta_3$-  (cyan shaded area) and
$\Delta_1$-contributions ($G_{\Delta_1}$, lower line, gray shaded area). The spin-resolved local density of states (LDOS) of the NN Co atom
and Mn atom are given in columns (b) and (c), respectively. For both (b) and (c) the LDOS spin-decomposition is performed
with respect to the global frame, with spin-up and spin-down LDOS presented in the upper and lower parts of each plot.
In both (b) and (c), the $\Delta_3$-contribution is indicated with solid lines, while for $\alpha=0^{\circ}$ and  $\alpha=180^{\circ}$ also the total LDOS is shown with dashed lines. The red triangles in (b) follow the
development of the dip in the LDOS of the NN Co atom as the angle $\alpha$ is varied.
In (c), the bonding and anti-bonding unoccupied Mn states are marked with "b" and "a", respectively.
For details see text.
}
\label{fig:T_vs_DOS}
\end{center}
\vspace*{-1cm}
\end{figure*}

In this section we perform a detailed analysis of the ballistic conductance $G(E_F)$ of the P$\alpha$ state at fixed
distance between Mn atoms of 4.5~bohr. At this distance, we calculate $G(E_F)$ as a function of angle $\alpha$
and present the results in Fig.~\ref{fig:rotate}(d). In this plot we observe that the conductance exhibits a very
non-trivial dependence on $\alpha$, originating mainly from the $\Delta_3$-orbitals 
($d_{xz,yz}$), while the $\Delta_1$ contribution ($s-d_{z^2}$) 
to the conductance, $G_{\Delta_1}$, remains almost perfectly constant. Surprisingly, the
$\Delta_3$-conductance almost vanishes for $\alpha$ of about 70$^{\circ}$, away from any high-symmetry spin
configuration in the junction, suggesting, that the dependence of the details of hybridization and electronic structure
on the angle between the Mn spins can be rather delicate. In order to analyze this dependence in more detail,
as a function of $\alpha$, we plot the energy-dependent conductance, $G(E)$, versus
the local densities of states (LDOS) of Mn and NN Co atoms resolved into spin-up and spin-down contributions with respect
to the global spin quantization $z$-axis, Fig.~\ref{fig:T_vs_DOS}. Mainly we focus on the $\Delta_3$-contribution
to the conductance and the LDOS, and
only
in the upper (P\up\up) and lower (P\down\down) panels
of Fig.~\ref{fig:T_vs_DOS} we show also the total LDOS of the atoms.

The conductance at a given energy $E$ depends on the presence of available states in the LDOS of the atoms at $E$, and
on the coupling between these states across the junction $-$ both of which depend on the orientation of the spins
with respect to each other. By looking at the LDOS of the atoms presented in Fig.~\ref{fig:T_vs_DOS} for $\alpha=0^{\circ}$ we can explain the absence of the $\Delta_4$ 
($d_{xy,x^2-y^2}$) 
contribution to the conductance:
the localized $\Delta_4$  states of the Co atoms, which can be seen as pronounced peaks in the LDOS 
marked with the dashed line in Fig.~\ref{fig:T_vs_DOS}, are positioned at about $-$2~eV for spin-up channel and
directly at the Fermi energy for spin-down channel, while the corresponding Mn $\Delta_4$ states are positioned below $-$2.5
and above $+$1~eV, prohibiting thus the hybridization between the Co and Mn orbitals of $\Delta_4$ symmetry
across the junction. 
Noticeably, 
the LDOS of both atoms for the up-spin in a wide region of energies around $E_F$ is
absent, leading to a negligible $\uparrow$-conductance. Here, it is important to remark, that the LDOS of the NN
Co atoms around the Fermi energy overall resembles quite well the LDOS of a Co atom in a 
Co monowire, see~e.g.~Fig.~13(a) in Ref.~\onlinecite{transportpaper}, or even of a Co atom
deposited on a noble-metal surfaces, see~e.g.~Ref.~\onlinecite{Nonas:2001}. This means that our results should be rather stable with respect to
the geometry of the Co leads, manifesting that the main influence on
the $\Delta_3$ conductance at $E_F$ would come from the hybridization of the Mn and NN Co states.

Turning now to the comparatively delocalized $\Delta_3$-states (solid line) on both Co and Mn atoms,
we observe  for P$\up\up$ (upper panels of Fig.~\ref{fig:T_vs_DOS})
that they hybridize directly
at the Fermi energy, which leads to a significant $\Delta_3$ contribution to the conductance. Specifically, while the
$\Delta_3^{\downarrow}$ subband of Co spreads from $-$1.8 to $+$1~eV, the  $\Delta_3$-down states of Mn atoms are
distinctly split into wide bonding ("b") states at the Fermi energy and narrow anti-bonding ("a") states at $+$1.8~eV.
Very importantly for the transport properties of the system, the hybridization
of the $\Delta_3^{\downarrow}$ band of Co with the $\Delta_3^{\downarrow}$ states of Mn is non-trivial.
(i) The $\Delta_3^{\downarrow}$ states of Co exhibit a dip at the position of the maximal density of bonding states
of Mn due to the fact that these Mn states are localized mainly in between the Mn atoms prohibiting strong
overlap with the Co states. (ii) The upper, antibonding, part of the Co $\Delta_3^{\downarrow}$ band hybridizes
stronger with the bonding states of Mn, since the antibonding states of Co atoms have a larger overlap with the
Mn orbitals, which results in a larger $\Delta_3$-conductance above $E_F$. (iii) Analogously, for energies below
$E_F$ the conductance is suppressed, since the bonding-like $\Delta_3^{\downarrow}$ Co states have smaller overlap
with the Mn bonding states.

Let us now follow the evolution of the electronic structure upon increasing the angle between the Mn spins.
Two trends in the LDOS can be clearly observed in Fig.~\ref{fig:T_vs_DOS}. Firstly, with increasing $\alpha$
the splitting between the bonding and antibonding Mn states decreases owing to the mixed spin character
of the states. At the angle of 90$^{\circ}$, when Mn spins are anti-parallel to each other, both types of states
transform into degenerate $\Delta_3$-orbitals of the "isolated" Mn atoms, since the hybridization between the
Mn states of the same spin is almost absent due to large exchange splitting. On the other hand, the dip in the
$\downarrow$-LDOS of the NN Co atoms follows the position of the bonding state of the Mn dimer, moving
from the Fermi energy at $\alpha=0^{\circ}$ to $+$0.2~eV for $\alpha=90^{\circ}$ (indicated by filled triangles
in Fig.~\ref{fig:T_vs_DOS}). Overall, such redistribution of the LDOS of the atoms combined with the effect of
decreasing LDOS of Mn atoms for spin-down channel at the Fermi energy when the angle $\alpha$ is varied,
results first in a decrease of the conductance at $E_F$ for $\alpha\approx 70^{\circ}$, followed by a consequent
increase with increasing angle.

When the
angle $\alpha$ increases further beyond 90$^{\circ}$, the bonding and anti-bonding Mn states
eventually acquire their initial splitting at $\alpha=180^{\circ}$ (P\down\down-state), when the Mn spins are collinear again.
Simultaneously, with increasing angle, we observe that the Mn states around the Fermi energy become
sharper, since the hybridization with the Co leads decreases as the Mn states become predominantly
spin-up in character. Interestingly, while for $90^{\circ} <\alpha < 120^{\circ}$ a large value
of the $\Delta_3$-conductance is due to a significant amount of delocalized Co and
bonding Mn states at the Fermi energy in the spin-down channel, for larger angles the value of
$G_{\Delta_3}(E_F)$ is due to a sharp resonant Co state in the spin-up channel at the Fermi energy,
coupled to a bonding Mn state at $E_F$. When further increasing $\alpha$ above 170$^{\circ}$,
this resonance becomes more localized and decoupled from the states in the leads, while the Mn LDOS
at the $E_F$ in the minority spin-channel vanishes, causing a sharp drop in the $\Delta_3$-conductance.
By looking at the total LDOS of the NN Co atom in the P\down\down-state we observe that it remains
basically unaffected, 
as compared to the P\up\up-configuration, while the Mn states become pronouncedly
decoupled from the states of the NN Co owing to the energetical mismatch for both spin channels.

\subsection{Fingerprints of non-collinear magnetic states of the junction in ballistic conductance
experiments}
\label{subsec:t-to-c_rev}

Finally, we investigate the evolution of the conductance and the magnetoresistance of different magnetic states of the junction 
within the contact regime mimicking a typical STM or break junction experiment. Here, we are partly motivated by the
fact that a non-trivial behavior of magnetoresistance when going from tunelling to contact has been recently
observed in STM experiments, see e.g.~Ref.~\onlinecite{ZieglerNJP2011}.
At a very large separation between the
leads (or, the tip and the sample in the STM language), owing to the FM coupling of the Mn atom
to the Co chain, one can imagine only two possible magnetic configurations $-$ P\up\up~and AP\up\down. The
conductance of these two magnetic states in the tunneling regime, arising mainly from the $s$-orbitals, is orders
of magnitude smaller than in the contact regime, for which the dependence of $G(E_F)$ on the distance
can be non-trivial due to the large contribution of the $d$-states.

\begin{figure}
\begin{center}
\centerline{\includegraphics[width=0.43\textwidth,angle=270]{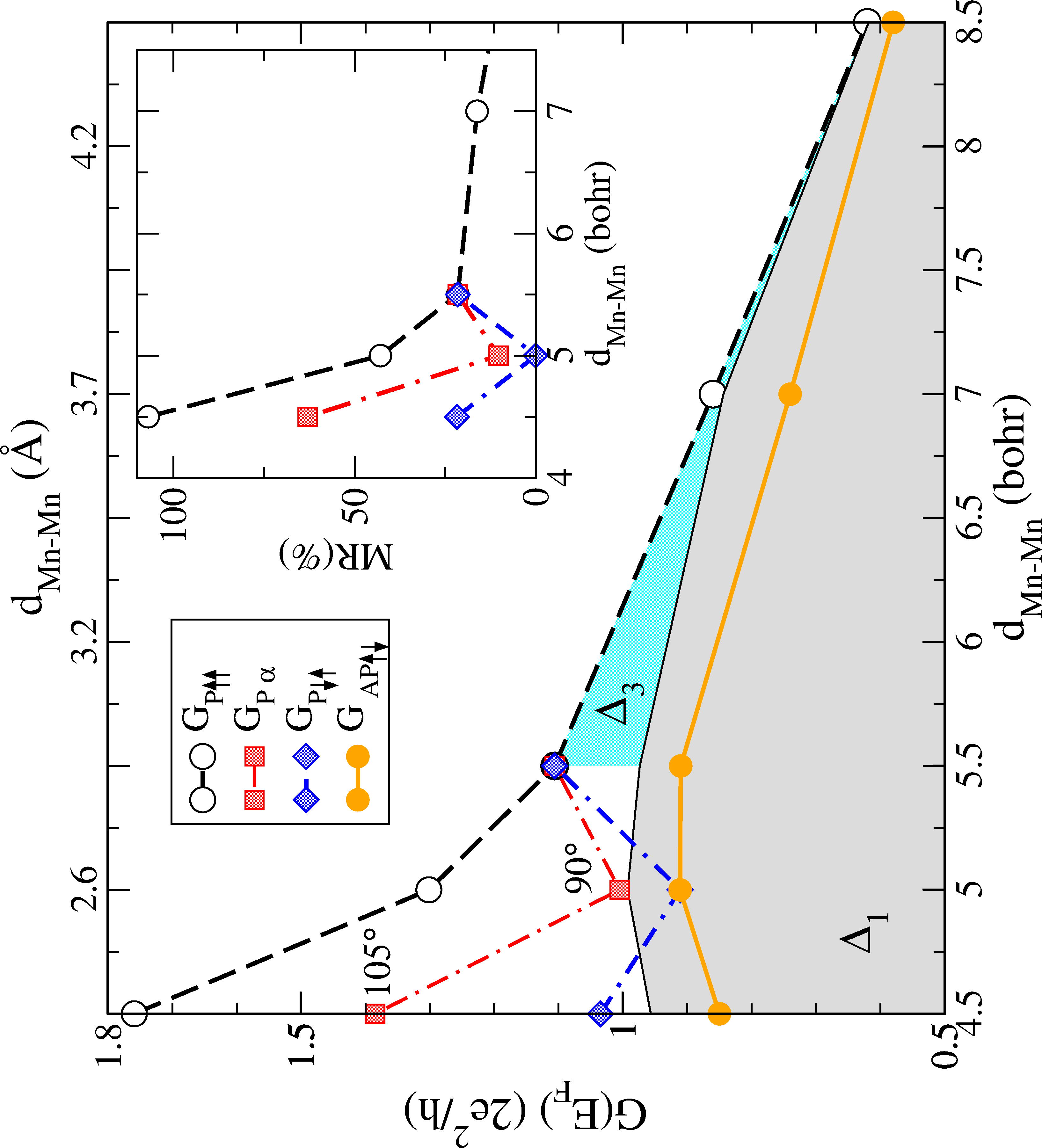}}
\caption{(color online)
Conductance at the Fermi energy of various possible magnetic states of the \cmmc~junction as a function of the
distance between the Mn atoms. Following magnetic states are considered: AP\down\up~(filled circles, solid line),
P\up\up~(open circles, dashed line), P\down\up~(diamonds, dash-dash-dotted line), and P$\alpha$ (squares,
dot-dashed line). The $\Delta_1$ contribution for the P\up\up-state is shown with a thin solid line and grey
shaded area, while the $\Delta_3$ part is shaded in cyan. For the P$\alpha$ states the state which is lowest in
energy among all possible angles $\alpha$ at a fixed distance is considered. In  the inset the values of the
magnetoresistance for different P-states are shown.
}
\label{fig:G}
\end{center}
\end{figure}

In the case
of the AP\up\down~configuration, the starting collinear arrangement of the spins will survive over the whole range
of the separation between the leads, since in the contact regime, when the Mn atoms are close to each other,
both exchange preferences of the Mn spins, that is, FM coupling to the NN Co spins and AFM coupling among each
other, are fulfilled. Small possible deviations from the collinear arrangement of the Mn spins, which can 
affect 
the details
of the distribution of the $\Delta_3$-states and their coupling to the leads, would not manifest in a conductance
measurement, owing to the 
antiparallel
 magnetizations of the leads, and corresponding complete dominance  of the
$\Delta_1$ channel for conductance at $E_F$ in this case, Fig.~\ref{fig:G}. As we can see from this figure, $G(E_F)$
lies in between 0.5 and 1.0$\,G_0$, when the distance between the Mn atoms is varied from 8.5 to 4.5~bohr.
This is very similar to the behavior of the conductance at the Fermi energy of pure AP Co leads without Mn atoms,
see e.g. Fig.~11(b) of Ref.~\onlinecite{transportpaper}.

Owing to the magnetic frustration of the Mn spins of the junction in the contact regime, for the P\up\up~initial
configuration, 
we consider the P$\alpha$ and P\down\up~states in addition to the P\up\up~state
when the Mn-Mn distance is relatively small. Here, as we have seen in the preceding section, the conductance
at the Fermi energy can be very strongly influenced by the details of hybridization between the $\Delta_3$-orbitals.
On the other hand, since very often transport measurements serve as the only experimental insight into the magnetic
structure of a system, it is very important to coin each of the possible magnetic states with a unique
fingerprint which can be related to the experimental data. Below, we suggest that indeed three distinct spin states
in a \cmmc~junction $-$ which can occur in an experiment due to various reasons such as structural details, temperature fluctuations,
external magnetic field, etc. $-$ lead to different transport signatures.

As already 
shown in Fig.~\ref{fig:stretch}(d), the conductance of the collinear P\up\up~state rapidly rises
towards a value of 1.8$\,G_0$ as the distance between the leads is decreased. Compared to other
possible magnetic configurations of the junction, $G_{{\rm P}\up\up}(E_F)$ is significantly larger in value, see
Fig.~\ref{fig:G}, because of the alignment of the 
minority spin $s$- and $d$-states 
of the Co electrodes and the Mn atoms at the
Fermi energy, 
which ideally favors perfect transmission. In contrast, the conductance of the
collinear P\down\up-state is significantly suppressed, reaching only 1.0$\,G_0$ at the separation
of 4.5~bohr, due to the large exchange splitting of the $\Delta_3$ states of the Mn atoms with
antiparallel 
spin moments which hinders the $\Delta_3$ conductance.
The conductance of the non-collinear ground-state P$\alpha$-state lies in between the values for
both limiting collinear configurations. In the close contact regime, at $d_{\rm{Mn-Mn}}$ of 4.5~bohr, the
conductance  of the P$\alpha$-state of 1.4$\,G_0$ is exactly in between the values of
$G_{{\rm P}\up\up}(E_F)$ and $G_{{\rm P}\down\up}(E_F)$. Clearly, the difference of 0.4$\,G_0$, 
stemming from
the variation in the $\Delta_3$-conductance with the spin state,  can be easily detected in experiment, 
allowing for a way to distinguish between different possible magnetic configurations. At the distance of 
5.0~bohr the ground state among the P$\alpha$ states is the P90$^{\circ}$ state, while at larger distances
above 5.5~bohr the system converges to a collinear configuration. The angle $\alpha$ in the lowest in energy P$\alpha$ 
state decreases smoothly with increasing the separation, and we speculate, that owing to the non-monotonous
behavior of the conductance as a function of $\alpha$, seen in Fig.~\ref{fig:rotate}(d), the conductance
as a function of $d_{\rm Mn-Mn}$, can exhibit several features similar to that at $d_{\rm Mn-Mn}$ of 5.0~bohr,
although we did not perform the calculations to support this statement owing to the required computational
effort. 

According to recent experiments,\cite{ZieglerNJP2011} the conductance of the junction with the parallel (P) and antiparallel (AP)
orientation of the lead's magnetization can be related to
each other via measuring the magnetoresistance (MR). From the values presented in Fig.~\ref{fig:G} we calculate the MR
of the junction, defined as:
\begin{equation}
\label{MR}
\mathrm{MR}=\frac{G_\mathrm{P}(E_F)-G_{\mathrm{AP}}(E_F)}{G_{\mathrm{AP}}(E_F)}\times 100\%,
\end{equation}
and present the MR as a function of separation between the electrodes in the inset of Fig.~\ref{fig:G}, where
we choose $G_{{\rm AP}\down\up}$ for $G_{\mathrm{AP}}(E_F)$, and values of $G_{{\rm P}\up\up}(E_F)$,
$G_{{\rm P}\down\up}(E_F)$ and $G_{{\rm P}\alpha}(E_F)$ for $G_\mathrm{P}(E_F)$.
The overall smaller AP\down\up~conductance as compared the P-configurations results in positive magnetoresistance
values. The MR curves as a function of the distance generally resemble those of the conductance, with the values
of the MR of 22, 62 and 105\% at the distance of 4.5~bohr for the P\down\up, P$\alpha$ and P\up\up-states, respectively.
Much more pronounced in the MR is the feature characteristic to the P\down\up~and P$\alpha$-configurations $-$ a dip around the
distance of 5.0~bohr, also present in the conductance curves. As can be seen from Fig.~\ref{fig:G}, at this distance,
the MR almost completely vanishes when the Mn spins exhibit a different from FM configuration. Overall, we
conclude, that the pronounced difference in the shape and magnitude of the MR curves can be also used in experiments
to shed light onto the complex magnetism in this type of systems.

\section{Summary}
\label{sec:Summary}

In this work we presented the realization of a first principles scheme for calculating the ballistic transport
properties of magnetically complex one-dimensional systems employing the technique of non-collinear Wannier functions.
We use the FLAPW method in order to calculate the electronic structure of the system with high accuracy
and use the Wannier functions to transfer
it to our transport calculations performed within the Landauer approach.
As spin-orbit interaction can be
naturally included into the consideration within this technique,~c.f.~Ref.~\onlinecite{transportpaper}, the method
introduced here
can be used to explore the rich field of transport phenomena in systems such as nano- or atomic-sized contacts,
break junctions, or STM experiments for which both effects, spin-orbit coupling and frustrated exchange interactions,
can be prominent.

As a first application of our approach, we consider the ballistic transport properties of a single-atom junction
formed by two semi-infinite Co electrodes with a single apex Mn atom. We study the conductance as a function of
the separation between the two Mn atoms from the tunneling to the contact regime taking into account the complex
magnetic interaction in the junction.
As we demonstrate, even such a simple setup allows to draw some general conclusions concerning the interplay of structure and
magnetism for the transport through such atomic-sized contacts which are in the focus of today's research. We analyze the ballistic conductance of the junction with lead magnetizations in parallel and antiparallel alignment.
We consider separately the tunneling (separation larger than about 5$\,$\AA) and contact (below 5$\,$\AA) regimes
of the junction, and we demonstrate that in the tunneling regime the conductance $G$ is solely coming from the overlap between
the $\Delta_1$ ($s$-$d_{z^2}$-orbitals) of the contacts. In this case the Mn spins prefer to order ferromagnetically
with respect to the magnetization of the leads. On the other hand, upon reaching the close contact regime (below
3.5$\,$\AA), when the magnitude of the conductance reaches 1$\,G_0$, the
hybridization between the $\Delta_3$ ($d_{xz},d_{yz}$-orbitals) states of the junction
starts to provide a sizable contribution to $G$.

In the close contact regime, when the hybridization between the Mn atoms is significant, Mn spins experience a
frustration due to the FM coupling with the leads and an AFM Mn-Mn coupling. The competition between the
two gives rise to a stable non-collinear solution which can be characterized by a tilting angle of the spins, $\alpha$.
General for this type of junction is the sensitivity of the $d$-orbital conductance on the angle $\alpha$, which is due to
a delicate interplay between the hybridization details of the Mn and Co states at the Fermi energy, as well as spin-asymmetry
in their distribution. This gives rise to a non-trivial $\alpha$-dependence of the conductance of the $d$-states. We
show that the complicated $\Delta_3$-channel conductance arising on the background of almost constant $\Delta_1$
contribution can be used in order to distinguish between different magnetic states of the contact via either a direct
conductance measurement, or via measuring the magnetoresistance, which, according to our calculations, can
vary in the contact regime between 20 and 100\%, depending on the spin arrangement.

Finally, we would like to comment on our approximation for the geometry of the junction we have assumed in
this work. Albeit being very simple, it allows to capture the key features which govern the transport properties
of the system, while keeping the computational burden reasonable. Namely, within this geometry: (i) the transition
from tunneling to contact can be naturally studied; (ii) the magnetic frustration of the spins in the junction, and
(iii) the delicate details of the hybridization of the adatom with the lead reservoirs are taken into account;
(iv) the sensitive dependence of the spin moments on the magnetic configuration in the
nano-contact is included into our considerations.  Of course, in order to achieve a quantitative agreement
of the calculated values to the experimentally measured ones in this type of junction beyond the major trends, all
details of the structure and structural reformation upon approaching should be ideally accounted for. Such
a challenging study lies, however, outside of the scope of the current work, and we leave it for future
studies.

\section{Acknowledgment}

We acknowledge helpful discussions with Stefan Bl\"ugel.
Funding by the DFG within the SFB677 is gratefully acknowledged.
S.H. thanks the DFG for financial support under HE3292/8-1. Y.M. and F.F. gratefully acknowledge the J\"ulich Supercomputing
Centre for computing time and funding under the HGF-YIG Programme VH-NG-513.


\begin{thebibliography}{48}
\expandafter\ifx\csname natexlab\endcsname\relax\def\natexlab#1{#1}\fi
\expandafter\ifx\csname bibnamefont\endcsname\relax
  \def\bibnamefont#1{#1}\fi
\expandafter\ifx\csname bibfnamefont\endcsname\relax
  \def\bibfnamefont#1{#1}\fi
\expandafter\ifx\csname citenamefont\endcsname\relax
  \def\citenamefont#1{#1}\fi
\expandafter\ifx\csname url\endcsname\relax
  \def\url#1{\texttt{#1}}\fi
\expandafter\ifx\csname urlprefix\endcsname\relax\def\urlprefix{URL }\fi
\providecommand{\bibinfo}[2]{#2}
\providecommand{\eprint}[2][]{\url{#2}}

\bibitem[{\citenamefont{Yanson et~al.}(1998)\citenamefont{Yanson, Bollinger,
  van~den Brom, Agrait, and van Ruitenbeek}}]{Yanson}
\bibinfo{author}{\bibfnamefont{A.~I.} \bibnamefont{Yanson}},
  \bibinfo{author}{\bibfnamefont{G.~R.} \bibnamefont{Bollinger}},
  \bibinfo{author}{\bibfnamefont{H.~E.} \bibnamefont{van~den Brom}},
  \bibinfo{author}{\bibfnamefont{N.}~\bibnamefont{Agrait}}, \bibnamefont{and}
  \bibinfo{author}{\bibfnamefont{J.~M.} \bibnamefont{van Ruitenbeek}},
  \bibinfo{journal}{Nature} \textbf{\bibinfo{volume}{395}},
  \bibinfo{pages}{783} (\bibinfo{year}{1998}).

\bibitem[{\citenamefont{Thiess et~al.}(2008)\citenamefont{Thiess, Mokrousov,
  Bl\"ugel, and Heinze}}]{ThiessNL2008}
\bibinfo{author}{\bibfnamefont{A.}~\bibnamefont{Thiess}},
  \bibinfo{author}{\bibfnamefont{Y.}~\bibnamefont{Mokrousov}},
  \bibinfo{author}{\bibfnamefont{S.}~\bibnamefont{Bl\"ugel}}, \bibnamefont{and}
  \bibinfo{author}{\bibfnamefont{S.}~\bibnamefont{Heinze}},
  \bibinfo{journal}{Nano Letters} \textbf{\bibinfo{volume}{8}},
  \bibinfo{pages}{2144} (\bibinfo{year}{2008}).

\bibitem[{\citenamefont{N\'eel et~al.}(2009)\citenamefont{N\'eel, Kr\"oger, and
  Berndt}}]{NeelPRL2009}
\bibinfo{author}{\bibfnamefont{N.}~\bibnamefont{N\'eel}},
  \bibinfo{author}{\bibfnamefont{J.}~\bibnamefont{Kr\"oger}}, \bibnamefont{and}
  \bibinfo{author}{\bibfnamefont{R.}~\bibnamefont{Berndt}},
  \bibinfo{journal}{Phys. Rev. Lett.} \textbf{\bibinfo{volume}{102}},
  \bibinfo{pages}{086805} (\bibinfo{year}{2009}).

\bibitem[{\citenamefont{Kr\"oger et~al.}(2008)\citenamefont{Kr\"oger, N\'eel,
  and Limot}}]{KroegerJPCM2008}
\bibinfo{author}{\bibfnamefont{J.}~\bibnamefont{Kr\"oger}},
  \bibinfo{author}{\bibfnamefont{N.}~\bibnamefont{N\'eel}}, \bibnamefont{and}
  \bibinfo{author}{\bibfnamefont{L.}~\bibnamefont{Limot}}, \bibinfo{journal}{J.
  Phys. Cond. Mat.} \textbf{\bibinfo{volume}{20}}, \bibinfo{pages}{223001}
  (\bibinfo{year}{2008}).

\bibitem[{\citenamefont{Tao et~al.}(2010)\citenamefont{Tao, Rungger, Sanvito,
  and Stepanyuk}}]{TaoPRB2010}
\bibinfo{author}{\bibfnamefont{K.}~\bibnamefont{Tao}},
  \bibinfo{author}{\bibfnamefont{I.}~\bibnamefont{Rungger}},
  \bibinfo{author}{\bibfnamefont{S.}~\bibnamefont{Sanvito}}, \bibnamefont{and}
  \bibinfo{author}{\bibfnamefont{V.~S.} \bibnamefont{Stepanyuk}},
  \bibinfo{journal}{Phys. Rev. B} \textbf{\bibinfo{volume}{82}},
  \bibinfo{pages}{085412} (\bibinfo{year}{2010}).

\bibitem[{\citenamefont{Chopra et~al.}(2005)\citenamefont{Chopra, Sullivan,
  Armstrong, and Hua}}]{ChopraNM2005}
\bibinfo{author}{\bibfnamefont{H.~D.} \bibnamefont{Chopra}},
  \bibinfo{author}{\bibfnamefont{M.~R.} \bibnamefont{Sullivan}},
  \bibinfo{author}{\bibfnamefont{J.~N.} \bibnamefont{Armstrong}},
  \bibnamefont{and} \bibinfo{author}{\bibfnamefont{S.~Z.} \bibnamefont{Hua}},
  \bibinfo{journal}{Nature Materials} \textbf{\bibinfo{volume}{4}},
  \bibinfo{pages}{832} (\bibinfo{year}{2005}).

\bibitem[{\citenamefont{Calvo et~al.}(2009{\natexlab{a}})\citenamefont{Calvo,
  Fern\'{a}ndez-Rossier, Palacios, Jacob, Natelson, and Untiedt}}]{CalvoN2009}
\bibinfo{author}{\bibfnamefont{M.}~\bibnamefont{Calvo}},
  \bibinfo{author}{\bibfnamefont{J.}~\bibnamefont{Fern\'{a}ndez-Rossier}},
  \bibinfo{author}{\bibfnamefont{J.}~\bibnamefont{Palacios}},
  \bibinfo{author}{\bibfnamefont{D.}~\bibnamefont{Jacob}},
  \bibinfo{author}{\bibfnamefont{D.}~\bibnamefont{Natelson}}, \bibnamefont{and}
  \bibinfo{author}{\bibfnamefont{C.}~\bibnamefont{Untiedt}},
  \bibinfo{journal}{Nature} \textbf{\bibinfo{volume}{458}},
  \bibinfo{pages}{1150} (\bibinfo{year}{2009}{\natexlab{a}}).

\bibitem[{\citenamefont{Ziegler et~al.}(2011)\citenamefont{Ziegler, N\'eel,
  Lazo, Ferriani, Heinze, Kr\"oger, and Berndt}}]{ZieglerNJP2011}
\bibinfo{author}{\bibfnamefont{M.}~\bibnamefont{Ziegler}},
  \bibinfo{author}{\bibfnamefont{N.}~\bibnamefont{N\'eel}},
  \bibinfo{author}{\bibfnamefont{C.}~\bibnamefont{Lazo}},
  \bibinfo{author}{\bibfnamefont{P.}~\bibnamefont{Ferriani}},
  \bibinfo{author}{\bibfnamefont{S.}~\bibnamefont{Heinze}},
  \bibinfo{author}{\bibfnamefont{J.}~\bibnamefont{Kr\"oger}}, \bibnamefont{and}
  \bibinfo{author}{\bibfnamefont{R.}~\bibnamefont{Berndt}},
  \bibinfo{journal}{New J. Phys.} \textbf{\bibinfo{volume}{17}},
  \bibinfo{pages}{085011} (\bibinfo{year}{2011}).

\bibitem[{\citenamefont{Schmaus et~al.}(2011)\citenamefont{Schmaus, Bagrets,
  Nahas, Yamada, Bork, Bowen, Beaurepaire, Evers, and
  Wulfhekel}}]{WulfhekelNN2011}
\bibinfo{author}{\bibfnamefont{S.}~\bibnamefont{Schmaus}},
  \bibinfo{author}{\bibfnamefont{A.}~\bibnamefont{Bagrets}},
  \bibinfo{author}{\bibfnamefont{Y.}~\bibnamefont{Nahas}},
  \bibinfo{author}{\bibfnamefont{T.~K.} \bibnamefont{Yamada}},
  \bibinfo{author}{\bibfnamefont{A.}~\bibnamefont{Bork}},
  \bibinfo{author}{\bibfnamefont{M.}~\bibnamefont{Bowen}},
  \bibinfo{author}{\bibfnamefont{E.}~\bibnamefont{Beaurepaire}},
  \bibinfo{author}{\bibfnamefont{F.}~\bibnamefont{Evers}}, \bibnamefont{and}
  \bibinfo{author}{\bibfnamefont{W.}~\bibnamefont{Wulfhekel}},
  \bibinfo{journal}{Nature Nanotechnology} \textbf{\bibinfo{volume}{6}},
  \bibinfo{pages}{185} (\bibinfo{year}{2011}).

\bibitem[{\citenamefont{Calvo et~al.}(2009{\natexlab{b}})\citenamefont{Calvo,
  Fern\'andez-Rossier, Palacios, Jacob, Natelson, and
  Untiedt}}]{CalvoNature2009}
\bibinfo{author}{\bibfnamefont{M.~R.} \bibnamefont{Calvo}},
  \bibinfo{author}{\bibfnamefont{J.}~\bibnamefont{Fern\'andez-Rossier}},
  \bibinfo{author}{\bibfnamefont{J.~J.} \bibnamefont{Palacios}},
  \bibinfo{author}{\bibfnamefont{D.}~\bibnamefont{Jacob}},
  \bibinfo{author}{\bibfnamefont{D.}~\bibnamefont{Natelson}}, \bibnamefont{and}
  \bibinfo{author}{\bibfnamefont{C.}~\bibnamefont{Untiedt}},
  \bibinfo{journal}{Nature} \textbf{\bibinfo{volume}{458}},
  \bibinfo{pages}{1150} (\bibinfo{year}{2009}{\natexlab{b}}).

\bibitem[{\citenamefont{Delin and Tosatti}(2003)}]{PhysRevB.68.144434}
\bibinfo{author}{\bibfnamefont{A.}~\bibnamefont{Delin}} \bibnamefont{and}
  \bibinfo{author}{\bibfnamefont{E.}~\bibnamefont{Tosatti}},
  \bibinfo{journal}{Phys. Rev. B} \textbf{\bibinfo{volume}{68}},
  \bibinfo{pages}{144434} (\bibinfo{year}{2003}).

\bibitem[{\citenamefont{Smogunov
  et~al.}(2008{\natexlab{a}})\citenamefont{Smogunov, Dal~Corso, Delin, Weht,
  and Tosatti}}]{SmogunovNN2008}
\bibinfo{author}{\bibfnamefont{A.}~\bibnamefont{Smogunov}},
  \bibinfo{author}{\bibfnamefont{A.}~\bibnamefont{Dal~Corso}},
  \bibinfo{author}{\bibfnamefont{A.}~\bibnamefont{Delin}},
  \bibinfo{author}{\bibfnamefont{R.}~\bibnamefont{Weht}}, \bibnamefont{and}
  \bibinfo{author}{\bibfnamefont{E.}~\bibnamefont{Tosatti}},
  \bibinfo{journal}{Nature Nanotechnology} \textbf{\bibinfo{volume}{3}},
  \bibinfo{pages}{22} (\bibinfo{year}{2008}{\natexlab{a}}).

\bibitem[{\citenamefont{Delin et~al.}(2004)\citenamefont{Delin, Tosatti, and
  Weht}}]{Delin2004}
\bibinfo{author}{\bibfnamefont{A.}~\bibnamefont{Delin}},
  \bibinfo{author}{\bibfnamefont{E.}~\bibnamefont{Tosatti}}, \bibnamefont{and}
  \bibinfo{author}{\bibfnamefont{R.}~\bibnamefont{Weht}},
  \bibinfo{journal}{Phys. Rev. Lett.} \textbf{\bibinfo{volume}{92}},
  \bibinfo{pages}{057201} (\bibinfo{year}{2004}).

\bibitem[{\citenamefont{Thiess et~al.}(2009)\citenamefont{Thiess, Mokrousov,
  Heinze, and Bl\"ugel}}]{Alex2009}
\bibinfo{author}{\bibfnamefont{A.}~\bibnamefont{Thiess}},
  \bibinfo{author}{\bibfnamefont{Y.}~\bibnamefont{Mokrousov}},
  \bibinfo{author}{\bibfnamefont{S.}~\bibnamefont{Heinze}}, \bibnamefont{and}
  \bibinfo{author}{\bibfnamefont{S.}~\bibnamefont{Bl\"ugel}},
  \bibinfo{journal}{Phys. Rev. Lett.} \textbf{\bibinfo{volume}{103}},
  \bibinfo{pages}{217201} (\bibinfo{year}{2009}).

\bibitem[{\citenamefont{Thiess et~al.}(2010)\citenamefont{Thiess, Mokrousov,
  and Heinze}}]{Mokrousov2010}
\bibinfo{author}{\bibfnamefont{A.}~\bibnamefont{Thiess}},
  \bibinfo{author}{\bibfnamefont{Y.}~\bibnamefont{Mokrousov}},
  \bibnamefont{and} \bibinfo{author}{\bibfnamefont{S.}~\bibnamefont{Heinze}},
  \bibinfo{journal}{Phys. Rev. B} \textbf{\bibinfo{volume}{81}},
  \bibinfo{pages}{054433} (\bibinfo{year}{2010}).

\bibitem[{\citenamefont{Hardrat et~al.}(2012)\citenamefont{Hardrat, Wang,
  Freimuth, Mokrousov, and Heinze}}]{transportpaper}
\bibinfo{author}{\bibfnamefont{B.}~\bibnamefont{Hardrat}},
  \bibinfo{author}{\bibfnamefont{N.-P.} \bibnamefont{Wang}},
  \bibinfo{author}{\bibfnamefont{F.}~\bibnamefont{Freimuth}},
  \bibinfo{author}{\bibfnamefont{Y.}~\bibnamefont{Mokrousov}},
  \bibnamefont{and} \bibinfo{author}{\bibfnamefont{S.}~\bibnamefont{Heinze}},
  \bibinfo{journal}{Phys. Rev. B} \textbf{\bibinfo{volume}{85}},
  \bibinfo{pages}{245412} (\bibinfo{year}{2012}).

\bibitem[{\citenamefont{Smogunov et~al.}(2004)\citenamefont{Smogunov,
  Dal~Corso, and Tosatti}}]{Smogunov:2004}
\bibinfo{author}{\bibfnamefont{A.}~\bibnamefont{Smogunov}},
  \bibinfo{author}{\bibfnamefont{A.}~\bibnamefont{Dal~Corso}},
  \bibnamefont{and} \bibinfo{author}{\bibfnamefont{E.}~\bibnamefont{Tosatti}},
  \bibinfo{journal}{Phys. Rev. B} \textbf{\bibinfo{volume}{70}},
  \bibinfo{pages}{045417} (\bibinfo{year}{2004}).

\bibitem[{\citenamefont{Smogunov et~al.}(2006)\citenamefont{Smogunov,
  Dal~Corso, and Tosatti}}]{Smogunov:2006}
\bibinfo{author}{\bibfnamefont{A.}~\bibnamefont{Smogunov}},
  \bibinfo{author}{\bibfnamefont{A.}~\bibnamefont{Dal~Corso}},
  \bibnamefont{and} \bibinfo{author}{\bibfnamefont{E.}~\bibnamefont{Tosatti}},
  \bibinfo{journal}{Phys. Rev. B} \textbf{\bibinfo{volume}{73}},
  \bibinfo{pages}{075418} (\bibinfo{year}{2006}).

\bibitem[{\citenamefont{Bagrets et~al.}(2004)\citenamefont{Bagrets,
  Papanikolaou, and Mertig}}]{Bagrets:2004}
\bibinfo{author}{\bibfnamefont{A.}~\bibnamefont{Bagrets}},
  \bibinfo{author}{\bibfnamefont{N.}~\bibnamefont{Papanikolaou}},
  \bibnamefont{and} \bibinfo{author}{\bibfnamefont{I.}~\bibnamefont{Mertig}},
  \bibinfo{journal}{Phys. Rev. B} \textbf{\bibinfo{volume}{70}},
  \bibinfo{pages}{064410} (\bibinfo{year}{2004}).

\bibitem[{\citenamefont{Bagrets et~al.}(2007)\citenamefont{Bagrets,
  Papanikolaou, and Mertig}}]{Bagrets:2007}
\bibinfo{author}{\bibfnamefont{A.}~\bibnamefont{Bagrets}},
  \bibinfo{author}{\bibfnamefont{N.}~\bibnamefont{Papanikolaou}},
  \bibnamefont{and} \bibinfo{author}{\bibfnamefont{I.}~\bibnamefont{Mertig}},
  \bibinfo{journal}{Phys. Rev. B} \textbf{\bibinfo{volume}{75}},
  \bibinfo{pages}{235448} (\bibinfo{year}{2007}).

\bibitem[{\citenamefont{Polok et~al.}(2011)\citenamefont{Polok, Fedorov,
  Bagrets, Zahn, and Mertig}}]{PolokPRB2011}
\bibinfo{author}{\bibfnamefont{M.}~\bibnamefont{Polok}},
  \bibinfo{author}{\bibfnamefont{D.~V.} \bibnamefont{Fedorov}},
  \bibinfo{author}{\bibfnamefont{A.}~\bibnamefont{Bagrets}},
  \bibinfo{author}{\bibfnamefont{P.}~\bibnamefont{Zahn}}, \bibnamefont{and}
  \bibinfo{author}{\bibfnamefont{I.}~\bibnamefont{Mertig}},
  \bibinfo{journal}{Phys. Rev. B} \textbf{\bibinfo{volume}{83}},
  \bibinfo{pages}{245426} (\bibinfo{year}{2011}).

\bibitem[{\citenamefont{Burton et~al.}(2006)\citenamefont{Burton, Sabirianov,
  Jaswal, Tsymbal, and Mryasov}}]{BurtonPRB2006}
\bibinfo{author}{\bibfnamefont{J.~D.} \bibnamefont{Burton}},
  \bibinfo{author}{\bibfnamefont{R.~F.} \bibnamefont{Sabirianov}},
  \bibinfo{author}{\bibfnamefont{S.~S.} \bibnamefont{Jaswal}},
  \bibinfo{author}{\bibfnamefont{E.~Y.} \bibnamefont{Tsymbal}},
  \bibnamefont{and} \bibinfo{author}{\bibfnamefont{O.~N.}
  \bibnamefont{Mryasov}}, \bibinfo{journal}{Phys. Rev. Lett.}
  \textbf{\bibinfo{volume}{97}}, \bibinfo{pages}{077204}
  (\bibinfo{year}{2006}).

\bibitem[{\citenamefont{Czerner et~al.}(2008)\citenamefont{Czerner, Yavorsky,
  and Mertig}}]{CzernerPRB2008}
\bibinfo{author}{\bibfnamefont{M.}~\bibnamefont{Czerner}},
  \bibinfo{author}{\bibfnamefont{B.~Y.} \bibnamefont{Yavorsky}},
  \bibnamefont{and} \bibinfo{author}{\bibfnamefont{I.}~\bibnamefont{Mertig}},
  \bibinfo{journal}{Phys. Rev. B} \textbf{\bibinfo{volume}{77}},
  \bibinfo{pages}{104411} (\bibinfo{year}{2008}).

\bibitem[{\citenamefont{Czerner et~al.}(2010)\citenamefont{Czerner, Yavorsky,
  and Mertig}}]{CzernerPSS2010}
\bibinfo{author}{\bibfnamefont{M.}~\bibnamefont{Czerner}},
  \bibinfo{author}{\bibfnamefont{B.~Y.} \bibnamefont{Yavorsky}},
  \bibnamefont{and} \bibinfo{author}{\bibfnamefont{I.}~\bibnamefont{Mertig}},
  \bibinfo{journal}{Phys. Status Solidi B} \textbf{\bibinfo{volume}{247}},
  \bibinfo{pages}{2594} (\bibinfo{year}{2010}).

\bibitem[{\citenamefont{Smogunov
  et~al.}(2008{\natexlab{b}})\citenamefont{Smogunov, Dal~Corso, and
  Tosatti}}]{Smogunov:2008}
\bibinfo{author}{\bibfnamefont{A.}~\bibnamefont{Smogunov}},
  \bibinfo{author}{\bibfnamefont{A.}~\bibnamefont{Dal~Corso}},
  \bibnamefont{and} \bibinfo{author}{\bibfnamefont{E.}~\bibnamefont{Tosatti}},
  \bibinfo{journal}{Phys. Rev. B} \textbf{\bibinfo{volume}{78}},
  \bibinfo{pages}{014423} (\bibinfo{year}{2008}{\natexlab{b}}).

\bibitem[{\citenamefont{Velev et~al.}(2005)\citenamefont{Velev, Sabirianov,
  Jaswal, and Tsymbal}}]{VelevPRL2005}
\bibinfo{author}{\bibfnamefont{J.}~\bibnamefont{Velev}},
  \bibinfo{author}{\bibfnamefont{R.~F.} \bibnamefont{Sabirianov}},
  \bibinfo{author}{\bibfnamefont{S.~S.} \bibnamefont{Jaswal}},
  \bibnamefont{and} \bibinfo{author}{\bibfnamefont{E.~Y.}
  \bibnamefont{Tsymbal}}, \bibinfo{journal}{Phys. Rev. Lett.}
  \textbf{\bibinfo{volume}{94}}, \bibinfo{pages}{127203}
  (\bibinfo{year}{2005}).

\bibitem[{\citenamefont{Bolotin et~al.}(2006)\citenamefont{Bolotin, Kuemmeth,
  and Ralph}}]{Bolotin2006}
\bibinfo{author}{\bibfnamefont{K.~I.} \bibnamefont{Bolotin}},
  \bibinfo{author}{\bibfnamefont{F.}~\bibnamefont{Kuemmeth}}, \bibnamefont{and}
  \bibinfo{author}{\bibfnamefont{D.~C.} \bibnamefont{Ralph}},
  \bibinfo{journal}{Phys. Rev. Lett.} \textbf{\bibinfo{volume}{97}},
  \bibinfo{pages}{127202} (\bibinfo{year}{2006}).

\bibitem[{\citenamefont{Burton et~al.}(2007)\citenamefont{Burton, Sabirianov,
  Velev, Mryasov, and Tsymbal}}]{BurtonPRB2007}
\bibinfo{author}{\bibfnamefont{J.~D.} \bibnamefont{Burton}},
  \bibinfo{author}{\bibfnamefont{R.~F.} \bibnamefont{Sabirianov}},
  \bibinfo{author}{\bibfnamefont{J.~P.} \bibnamefont{Velev}},
  \bibinfo{author}{\bibfnamefont{O.~N.} \bibnamefont{Mryasov}},
  \bibnamefont{and} \bibinfo{author}{\bibfnamefont{E.~Y.}
  \bibnamefont{Tsymbal}}, \bibinfo{journal}{Phys. Rev. B}
  \textbf{\bibinfo{volume}{76}}, \bibinfo{pages}{144430}
  (\bibinfo{year}{2007}).

\bibitem[{\citenamefont{Tao et~al.}(2009)\citenamefont{Tao, Stepanyuk, Hergert,
  Rungger, Sanvito, and Bruno}}]{TaoPRL2009}
\bibinfo{author}{\bibfnamefont{K.}~\bibnamefont{Tao}},
  \bibinfo{author}{\bibfnamefont{V.~S.} \bibnamefont{Stepanyuk}},
  \bibinfo{author}{\bibfnamefont{W.}~\bibnamefont{Hergert}},
  \bibinfo{author}{\bibfnamefont{I.}~\bibnamefont{Rungger}},
  \bibinfo{author}{\bibfnamefont{S.}~\bibnamefont{Sanvito}}, \bibnamefont{and}
  \bibinfo{author}{\bibfnamefont{P.}~\bibnamefont{Bruno}},
  \bibinfo{journal}{Phys. Rev. Lett.} \textbf{\bibinfo{volume}{103}},
  \bibinfo{pages}{057202} (\bibinfo{year}{2009}).

\bibitem[{\citenamefont{Serrate et~al.}(2010)\citenamefont{Serrate, Ferriani,
  Yoshida, Hla, Menzel, von Bergmann, Heinze, Kubetzka, and
  Wiesendanger}}]{SerrateNN2010}
\bibinfo{author}{\bibfnamefont{D.}~\bibnamefont{Serrate}},
  \bibinfo{author}{\bibfnamefont{P.}~\bibnamefont{Ferriani}},
  \bibinfo{author}{\bibfnamefont{Y.}~\bibnamefont{Yoshida}},
  \bibinfo{author}{\bibfnamefont{S.-W.} \bibnamefont{Hla}},
  \bibinfo{author}{\bibfnamefont{M.}~\bibnamefont{Menzel}},
  \bibinfo{author}{\bibfnamefont{K.}~\bibnamefont{von Bergmann}},
  \bibinfo{author}{\bibfnamefont{S.}~\bibnamefont{Heinze}},
  \bibinfo{author}{\bibfnamefont{A.}~\bibnamefont{Kubetzka}}, \bibnamefont{and}
  \bibinfo{author}{\bibfnamefont{R.}~\bibnamefont{Wiesendanger}},
  \bibinfo{journal}{Nature Nanotechnology} \textbf{\bibinfo{volume}{5}},
  \bibinfo{pages}{350} (\bibinfo{year}{2010}).

\bibitem[{\citenamefont{Menzel et~al.}(2012)\citenamefont{Menzel, Mokrousov,
  Wieser, Bickel, Vedmedenko, Bl\"ugel, Heinze, von Bergmann, Kubetzka, and
  Wiesendanger}}]{Menzel:2012}
\bibinfo{author}{\bibfnamefont{M.}~\bibnamefont{Menzel}},
  \bibinfo{author}{\bibfnamefont{Y.}~\bibnamefont{Mokrousov}},
  \bibinfo{author}{\bibfnamefont{R.}~\bibnamefont{Wieser}},
  \bibinfo{author}{\bibfnamefont{J.~E.} \bibnamefont{Bickel}},
  \bibinfo{author}{\bibfnamefont{E.}~\bibnamefont{Vedmedenko}},
  \bibinfo{author}{\bibfnamefont{S.}~\bibnamefont{Bl\"ugel}},
  \bibinfo{author}{\bibfnamefont{S.}~\bibnamefont{Heinze}},
  \bibinfo{author}{\bibfnamefont{K.}~\bibnamefont{von Bergmann}},
  \bibinfo{author}{\bibfnamefont{A.}~\bibnamefont{Kubetzka}}, \bibnamefont{and}
  \bibinfo{author}{\bibfnamefont{R.}~\bibnamefont{Wiesendanger}},
  \bibinfo{journal}{Phys. Rev. Lett.} \textbf{\bibinfo{volume}{108}},
  \bibinfo{pages}{197204} (\bibinfo{year}{2012}).

\bibitem[{\citenamefont{Hohenberg and Kohn}(1964)}]{PhysRev.136.B864}
\bibinfo{author}{\bibfnamefont{P.}~\bibnamefont{Hohenberg}} \bibnamefont{and}
  \bibinfo{author}{\bibfnamefont{W.}~\bibnamefont{Kohn}},
  \bibinfo{journal}{Phys. Rev.} \textbf{\bibinfo{volume}{136}},
  \bibinfo{pages}{B864} (\bibinfo{year}{1964}).

\bibitem[{\citenamefont{Kurz et~al.}(2004)\citenamefont{Kurz, F\"orster,
  Nordstr\"om, Bihlmayer, and Bl\"ugel}}]{KurznocoFLAPW}
\bibinfo{author}{\bibfnamefont{P.}~\bibnamefont{Kurz}},
  \bibinfo{author}{\bibfnamefont{F.}~\bibnamefont{F\"orster}},
  \bibinfo{author}{\bibfnamefont{L.}~\bibnamefont{Nordstr\"om}},
  \bibinfo{author}{\bibfnamefont{G.}~\bibnamefont{Bihlmayer}},
  \bibnamefont{and} \bibinfo{author}{\bibfnamefont{S.}~\bibnamefont{Bl\"ugel}},
  \bibinfo{journal}{Phys. Rev. B} \textbf{\bibinfo{volume}{69}},
  \bibinfo{pages}{024415} (\bibinfo{year}{2004}).

\bibitem[{\citenamefont{Wannier}(1937)}]{WannierPR1937}
\bibinfo{author}{\bibfnamefont{G.~H.} \bibnamefont{Wannier}},
  \bibinfo{journal}{Phys. Rev.} \textbf{\bibinfo{volume}{52}},
  \bibinfo{pages}{191} (\bibinfo{year}{1937}).

\bibitem[{\citenamefont{Marzari and Vanderbilt}(1997)}]{MarzariPRB1997}
\bibinfo{author}{\bibfnamefont{N.}~\bibnamefont{Marzari}} \bibnamefont{and}
  \bibinfo{author}{\bibfnamefont{D.}~\bibnamefont{Vanderbilt}},
  \bibinfo{journal}{Phys. Rev. B} \textbf{\bibinfo{volume}{56}},
  \bibinfo{pages}{12847} (\bibinfo{year}{1997}).

\bibitem[{\citenamefont{Freimuth et~al.}(2008)\citenamefont{Freimuth,
  Mokrousov, Wortmann, Heinze, and Bl\"ugel}}]{FreimuthPRB2008}
\bibinfo{author}{\bibfnamefont{F.}~\bibnamefont{Freimuth}},
  \bibinfo{author}{\bibfnamefont{Y.}~\bibnamefont{Mokrousov}},
  \bibinfo{author}{\bibfnamefont{D.}~\bibnamefont{Wortmann}},
  \bibinfo{author}{\bibfnamefont{S.}~\bibnamefont{Heinze}}, \bibnamefont{and}
  \bibinfo{author}{\bibfnamefont{S.}~\bibnamefont{Bl\"ugel}},
  \bibinfo{journal}{Phys. Rev. B} \textbf{\bibinfo{volume}{78}},
  \bibinfo{pages}{035120} (\bibinfo{year}{2008}).

\bibitem[{fle()}]{fleur}
\urlprefix\url{www.flapw.de}.

\bibitem[{\citenamefont{Krakauer et~al.}(1979)\citenamefont{Krakauer,
  Posternak, and Freeman}}]{FilmFLAPW}
\bibinfo{author}{\bibfnamefont{H.}~\bibnamefont{Krakauer}},
  \bibinfo{author}{\bibfnamefont{M.}~\bibnamefont{Posternak}},
  \bibnamefont{and} \bibinfo{author}{\bibfnamefont{A.~J.}
  \bibnamefont{Freeman}}, \bibinfo{journal}{Phys. Rev. B}
  \textbf{\bibinfo{volume}{19}}, \bibinfo{pages}{1706} (\bibinfo{year}{1979}).

\bibitem[{\citenamefont{Mokrousov et~al.}(2005)\citenamefont{Mokrousov,
  Bihlmayer, and Bl\"ugel}}]{Mokrousov:2005}
\bibinfo{author}{\bibfnamefont{Y.}~\bibnamefont{Mokrousov}},
  \bibinfo{author}{\bibfnamefont{G.}~\bibnamefont{Bihlmayer}},
  \bibnamefont{and} \bibinfo{author}{\bibfnamefont{S.}~\bibnamefont{Bl\"ugel}},
  \bibinfo{journal}{Phys. Rev. B} \textbf{\bibinfo{volume}{72}},
  \bibinfo{pages}{045402} (\bibinfo{year}{2005}).

\bibitem[{\citenamefont{Kizuka}(2008)}]{KizukaPRB2008}
\bibinfo{author}{\bibfnamefont{T.}~\bibnamefont{Kizuka}},
  \bibinfo{journal}{Phys. Rev. B} \textbf{\bibinfo{volume}{77}},
  \bibinfo{pages}{155401} (\bibinfo{year}{2008}).

\bibitem[{\citenamefont{Zhang and Yang}(1998)}]{rpbe}
\bibinfo{author}{\bibfnamefont{Y.}~\bibnamefont{Zhang}} \bibnamefont{and}
  \bibinfo{author}{\bibfnamefont{W.}~\bibnamefont{Yang}},
  \bibinfo{journal}{Phys. Rev. Lett.} \textbf{\bibinfo{volume}{80}},
  \bibinfo{pages}{890} (\bibinfo{year}{1998}).

\bibitem[{\citenamefont{Birkenheuer and Izotov}(2005)}]{disentanglement}
\bibinfo{author}{\bibfnamefont{U.}~\bibnamefont{Birkenheuer}} \bibnamefont{and}
  \bibinfo{author}{\bibfnamefont{D.}~\bibnamefont{Izotov}},
  \bibinfo{journal}{Phys. Rev. B} \textbf{\bibinfo{volume}{71}},
  \bibinfo{pages}{125116} (\bibinfo{year}{2005}).

\bibitem[{\citenamefont{Mokrousov et~al.}(2007)\citenamefont{Mokrousov,
  Bihlmayer, Bl\"ugel, and Heinze}}]{PhysRevB.75.104413}
\bibinfo{author}{\bibfnamefont{Y.}~\bibnamefont{Mokrousov}},
  \bibinfo{author}{\bibfnamefont{G.}~\bibnamefont{Bihlmayer}},
  \bibinfo{author}{\bibfnamefont{S.}~\bibnamefont{Bl\"ugel}}, \bibnamefont{and}
  \bibinfo{author}{\bibfnamefont{S.}~\bibnamefont{Heinze}},
  \bibinfo{journal}{Phys. Rev. B} \textbf{\bibinfo{volume}{75}},
  \bibinfo{pages}{104413} (\bibinfo{year}{2007}).

\bibitem[{\citenamefont{Schubert et~al.}(2011)\citenamefont{Schubert,
  Mokrousov, Ferriani, and Heinze}}]{franziska}
\bibinfo{author}{\bibfnamefont{F.}~\bibnamefont{Schubert}},
  \bibinfo{author}{\bibfnamefont{Y.}~\bibnamefont{Mokrousov}},
  \bibinfo{author}{\bibfnamefont{P.}~\bibnamefont{Ferriani}}, \bibnamefont{and}
  \bibinfo{author}{\bibfnamefont{S.}~\bibnamefont{Heinze}},
  \bibinfo{journal}{Phys. Rev. B} \textbf{\bibinfo{volume}{83}},
  \bibinfo{pages}{165442} (\bibinfo{year}{2011}).

\bibitem[{\citenamefont{Zelen\'y et~al.}(2009)\citenamefont{Zelen\'y,
  \ifmmode~\check{S}\else \v{S}\fi{}ob, and Hafner}}]{zeleny}
\bibinfo{author}{\bibfnamefont{M.}~\bibnamefont{Zelen\'y}},
  \bibinfo{author}{\bibfnamefont{M.}~\bibnamefont{\ifmmode~\check{S}\else
  \v{S}\fi{}ob}}, \bibnamefont{and}
  \bibinfo{author}{\bibfnamefont{J.}~\bibnamefont{Hafner}},
  \bibinfo{journal}{Phys. Rev. B} \textbf{\bibinfo{volume}{80}},
  \bibinfo{pages}{144414} (\bibinfo{year}{2009}).

\bibitem[{\citenamefont{Ataca et~al.}(2008)\citenamefont{Ataca, Cahangirov,
  Durgun, Jang, and Ciraci}}]{ataca}
\bibinfo{author}{\bibfnamefont{C.}~\bibnamefont{Ataca}},
  \bibinfo{author}{\bibfnamefont{S.}~\bibnamefont{Cahangirov}},
  \bibinfo{author}{\bibfnamefont{E.}~\bibnamefont{Durgun}},
  \bibinfo{author}{\bibfnamefont{Y.-R.} \bibnamefont{Jang}}, \bibnamefont{and}
  \bibinfo{author}{\bibfnamefont{S.}~\bibnamefont{Ciraci}},
  \bibinfo{journal}{Phys. Rev. B} \textbf{\bibinfo{volume}{77}},
  \bibinfo{pages}{214413} (\bibinfo{year}{2008}).

\bibitem[{\citenamefont{Gunnarsson}(1976)}]{Gunnarson_ICo}
\bibinfo{author}{\bibfnamefont{O.}~\bibnamefont{Gunnarsson}},
  \bibinfo{journal}{Journal of Physics F: Metal Physics}
  \textbf{\bibinfo{volume}{6}}, \bibinfo{pages}{587} (\bibinfo{year}{1976}).

\bibitem[{\citenamefont{Nonas et~al.}(2001)\citenamefont{Nonas, Cabria, Zeller,
  Dederichs, Huhne, and Ebert}}]{Nonas:2001}
\bibinfo{author}{\bibfnamefont{B.}~\bibnamefont{Nonas}},
  \bibinfo{author}{\bibfnamefont{I.}~\bibnamefont{Cabria}},
  \bibinfo{author}{\bibfnamefont{R.}~\bibnamefont{Zeller}},
  \bibinfo{author}{\bibfnamefont{P.~H.} \bibnamefont{Dederichs}},
  \bibinfo{author}{\bibfnamefont{T.}~\bibnamefont{Huhne}}, \bibnamefont{and}
  \bibinfo{author}{\bibfnamefont{H.}~\bibnamefont{Ebert}},
  \bibinfo{journal}{Phys. Rev. Lett.} \textbf{\bibinfo{volume}{86}},
  \bibinfo{pages}{2146} (\bibinfo{year}{2001}).

\end{thebibliography}
\end{document}